\documentclass[%
 aps,
 prd,
 amsmath,amssymb,
 reprint,%
superscriptaddress
]{revtex4-1}

\usepackage{graphicx}% Include figure files
\usepackage{grffile}
\usepackage{dcolumn}% Align table columns on decimal point
\usepackage{bm}% bold math
\usepackage{mathrsfs}
\usepackage{color}
\begin{document}

\title[regularization study]{Bias-free parameter estimation depending on sky region for targeted gravitational wave}% Force line breaks with \\

\author{Kenji Ono}%
 \email{kenji@icrr.u-tokyo.ac.jp}
\affiliation{
Institute for Cosmic Ray Research (ICRR), The university of Tokyo, 5-1-5 Kashiwanoha, Kashiwa, Chiba 277-8582, Japan%\\This line break forced with \textbackslash\textbackslash
}%
\affiliation{Department of Physics, Graduate School of Science, The university of Tokyo, Tokyo 133-0033, Japan
%\\This line break forced% with \\
}%

\author{Kazuhiro Hayama}
\affiliation{KAGRA Observatory, Institute for Cosmic Ray Research, The university of Tokyo, 238 Higashi Mozumi, Kamioka, Hida, Gifu 506-1205, Japan
%\\This line break forced% with \\
}%

\affiliation{Department of Applied Physics, Fukuoka University, Jonan, Nanakuma, Fukuoka 814-0180, Japan
%\\This line break forced% with \\
}%
\date{\today}% It is always \today, today,

\begin{abstract}
Detection of gravitational waves(GW) involves using the network of GW telescopes to observe a large sky region. However, owing to the arrangement of the GW telescopes, even with aLIGO-aVirgo-KAGRA network,parameter estimation accuracy deteriorates depending on the sky region of the GW source due to the ill-posed nature of the inverse operator. A regularization-based method is proposed herein to improve parameter estimation accuracy. Although conventional regularization methods cannot optimize all regulator parameters completely, the proposed method archives full optimization by applying an a-posteriori parameter choice rule to determine regulator parameters. We demonstrate that the proposed method can shrink the credible regions of inclination vs luminosity distance and polarization vs initial phase significantly in the sky wherein the accuracy of the amplitude parameters of a GW has been deteriorated. The proposed method suppresses the systematic error of a GW depending on the sky region and allows us investigating the cosmological information more precisely.

\end{abstract}

% \keywords{Suggested keywords}%Use showkeys class option if keyword
\maketitle

\section{\label{sec:intro}Introduction}
Gravitational waves(GW) have been detected using extremely large laser interferometers. i.e., GW telescopes. Km-scale GW telescopes, such as LIGO\cite{0034-4885-72-7-076901} and Virgo\cite{0264-9381-32-2-024001} have been constructed and the KAGRA\cite{PhysRevD.88.043007} is currently under construction in Japan. LIGO Scientific collaboration and Virgo collaboration has successfully detected a GW directly from a compact binary coalescence (CBC) comprising two black holes\cite{PhysRevLett.116.061102}. Astronomers and physicists expect that GW astronomy will reveal phenomena that have not been previously clarified via electromagnetic astronomy.

In future, it is expected that more precise estimation of GW waveforms or parameters will be achieved using a network of GW telescopes\cite{PhysRevD.88.062001}, i.e., using multiple telescopes simultaneously to detect a GW. With a network of GW telescopes, the SNR of GW detection can be increased and the independent mode of a GW can be determined. An analysis method based on Bayesian statistics has been proposed\cite{PhysRevD.83.084002} to estimate GW waveforms detected by a network of GW telescopes. To estimate the amplitude parameters of a GW, maximizing the likelihood of the output of GW telescopes and the GW model are equivalent to solving an inverse problem whose inverse operator considers the parameters of antenna-beam pattern functions and detecting the SNR of a GW\cite{PhysRevD.83.084002}. However, the solution of an inverse problem is unstable owing to the rank deficiency of the inverse operator. The instability of the solution, i.e., an ill-posed problem, makes it impossible to distinguish the independent modes of GW due to the degeneration of these modes. In order to avoid the ill-posed problem, certain solutions are suggested for detecting a GW by the network of GW telescopes, especially for burst search of GW. For instance, Rakhmanov formulated the Maximum likelihood method with Tikhonov regularization\cite{0264-9381-23-19-S05} and Mohanty propose to constract a regulator by a variability of the SNR as the source is displaced on the sky\cite{0264-9381-23-15-001}.

However, conventional regularization methods for a coherent search focus on reducing the amplified noise to the theoretical limit and ignore the fact that estimated GW parameters can exceed the value range of the actual parameters because the regulator adds bias noise\cite{engl1996regularization}. This is a significant problem because it is highly likely that the actual GW parameters are outside the credible region estimated using a regularization method. The bias noise introduced by a regulator can cause inaccurate estimation of parameter values. To implement precise estimation of GW parameters, a regularization method is required to reduce the amplified noise as small as possible while maintaining the bias noise such that the estimation points of the GW parameters are not affected. Thus, we propose a method to optimize regulator parameters for minimizing the influence of amplitude parameters amplified by the ill-posed inverse operator in the analysis of a targeted CBC search. The proposed method attempts to minimize the residual of the amplitude parameters expressed by the sum of amplified and bias noise. The residual of the amplitude parameters of a GW, which is expressed by the norm of the difference between the actual amplitudes and the estimated amplitudes evaluated using regularized data analysis, must be minimized to obtain all of the optimized parameters of a regulator. However, the actual GW amplitudes cannot be predetermined. Thus, the optimized regulator parameters are selected based on an a-posteriori parameter choice rule\cite{engl1996regularization}. The estimated amplitude allows us to evaluate the value of the norm. Then, it becomes possible to determine full optimized parameters of a regulator. We implement data analysis with the optimized regulator to improve the accuracy of the amplitude parameters of a GW.

The remainder of this paper is organized as follows. In Section.\ref{server-thesis-sec:coher-match-filt}, we describe the fundamental formulation of a targeted coherent CBC search. We explain how the optimized regulator in the coherent search method is determined using an a-posteirori parameter choice rule in Section.\ref{ken-short_thesis-sec:regul-inverse-probl}. In Section.\ref{ken-short_thesis-sec:analysis}, we review the regularized data analysis algorithm and describe MultiNest\cite{2009MNRAS.398.1601F}\cite{2013arXiv1306.2144F}, which is Bayesian inference tool. In addition, we present the results of regularized data analysis and describe the behavior of the reduction of the amplified noise. Conclusions are given in Section.\ref{server-thesis-sec:conclusion}.

\section{coherent matched filtering for GWs from CBC}\label{server-thesis-sec:coher-match-filt}
In this section, we describe a coherent search method\cite{PhysRevD.83.084002} with a network of GW telescopes. For simplification, we suppose the terms of the GW phase have Newtonian approximation because we focus on the improvement of the accuracy the parameters of GW amplitude which include the inclination of a binary system, the luminosity distance, the polarization and the initial phase. The accuracy of these parameters is mainly improved by application of the regularization method to the inverse operator into a likelihood. It will be discussed Section.\ref{ken-short_thesis-sec:regul-inverse-probl} in details.

According to \cite{PhysRevD.58.063001}, a response function
$h(t)$ can be written in the time domain as
\begin{equation}\label{ken-thesis-eq:11}
 h(t) = F_{+}(t)h_{+}(t) + F_{\times}(t)h_{\times}(t),
\end{equation}
where
$F_{+}(t)$ and
$F_{\times}(t)$ are called beam-pattern functions. These functions depend the geometry of a GW telescope and the local sidereal time(LST) at a GW telescope:
\begin{eqnarray}
 F_{+}(t) &=& \sin\zeta \left[F_{+,0}(t)\cos 2\Phi + F_{\times,0}(t)\sin 2\Phi\right],\label{ken-thesis-eq:12}\\
 F_{\times}(t) &=& \sin\zeta \left[-F_{\times,0}(t)\cos 2\Phi + F_{+,0}(t)\sin 2\Phi\right],\label{ken-thesis-eq:13}
\end{eqnarray}
where
$\zeta$ is the angle between the arms of a GW telescope, and
$\Phi$ is the polarization angle.
$F_{+,0}(t)$ and
$F_{\times,0}(t)$ are the functions when the polarization angle is set to 0\cite{PhysRevD.58.063001}.

According to \cite{0264-9381-24-23-001}, the response function of a GW telescope can be decomposed into 4 independent terms as follows:
\begin{widetext}
\begin{eqnarray}\label{ken-thesis-eq:16}
 h(t) &=& A^{0}F_{+,0}(t_{\mathrm{coal}})h_{0}(t) + A^{1}F_{\times,0}(t_{\mathrm{coal}})h_{0}(t) + A^{2}F_{+,0}(t_{\mathrm{coal}})h_{\pi/2}(t) + A^{3}F_{\times,0}(t_{\mathrm{coal}})h_{\pi/2}(t)\nonumber\\
 &\equiv& A^{\mu}h_{\mu}(t).
\end{eqnarray}
\end{widetext}
The meaning of the parameters in the function are as follows: the mutually independent phase
$h_{0}$ and
$h_{\pi/2}$ are given by
\begin{eqnarray}
 h_{0}(t) &=& \frac{1}{c}\left(\frac{G M_{c}}{c^{2}}\right)^{5/4}\left(\frac{5}{c \tau}\right)^{1/4}\cos\Psi (M_{c}, \tau),\label{ken-thesis-eq:17}\\
 h_{\pi/2}(t) &=& \frac{1}{c}\left(\frac{G M_{c}}{c^{2}}\right)^{5/4}\left(\frac{5}{c \tau}\right)^{1/4}\sin\Psi (M_{c}, \tau),
\end{eqnarray}
where 
\begin{equation}\label{ken-short_thesis-eq:4}
 \Psi (M_{c}, \tau) = -2\left(\frac{5GM_{c}}{c^{3}}\right)^{-5/8}\tau^{5/8},
\end{equation}
$G$ is the gravitational constant,
$c$ is the light speed,
$M_{c}$ is the chirp mass defined by
\begin{equation}\label{ken-thesis-eq:5}
 M_{c} = \frac{(m_{1}m_{2})^{3/5}}{(m_{1}+m_{2})^{1/5}},
\end{equation}
where the
$m_{1}$ and
$m_{2}$ are the masses of the stars of a binary system, and
$\tau$ is defined by
\begin{equation}\label{ken-thesis-eq:3}
 \tau = t_{\mathrm{coal}}-t,
\end{equation}
where
$t$ is the observer time and
$t_{\mathrm{coal}}$ is the observer time when a compact binary system coalesces. the GW amplitudes
$A^{\mu}\ (\mu=0,1,2,3)$ are given by
\begin{eqnarray*}
 A^{0} &=& \frac{c}{r}\left[\frac{1+\cos^{2}\iota}{2}\cos 2\Phi \cos\phi_{\mathrm{coal}} \allowbreak- \cos\iota\sin 2\Phi\sin\phi_{\mathrm{coal}}\right],\label{ken-thesais-eq:18}\\
 A^{1} &=& \frac{c}{r}\left[\frac{1+\cos^{2}\iota}{2}\sin 2\Phi \cos\phi_{\mathrm{coal}} + \cos\iota\cos 2\Phi\sin\phi_{\mathrm{coal}}\right],\label{ken-thesis-eq:19}\\
 A^{2} &=& \frac{c}{r}\left[-\frac{1+\cos^{2}\iota}{2}\cos 2\Phi \sin\phi_{\mathrm{coal}} - \cos\iota\sin 2\Phi\cos\phi_{\mathrm{coal}}\right],\label{ken-thesis-eq:20}\\
 A^{3} &=& \frac{c}{r}\left[-\frac{1+\cos^{2}\iota}{2}\sin 2\Phi \sin\phi_{\mathrm{coal}} + \cos\iota\cos 2\Phi\cos\phi_{\mathrm{coal}}\right],\label{ken-thesis-eq:21}
\end{eqnarray*}
where
$\iota$ is the inclination of a binary system, and an initial phase
$\phi_{\mathrm{coal}}$ is the phase at
$t_{\mathrm{coal}}$. Note that we suppose the beam-pattern functions are constants during detecting a GW by the use of the network of GW telescopes. The approximation is reasonable because the time over detecting a GW is too short to change the value of beam-pattern functions.

Next, we describe the formalism of a coherent search method for the GW from a CBC. The index
$X$ indicates the X-th GW telescope belonging to the network of GW telescopes:
\begin{equation}\label{ken-thesis-eq:25}
 h^{X}(t)=A^{\mu}h^{X}_{\mu}(t).
\end{equation}

One fundamental detection statistics of a coherent search method in this thesis is the matched filtering; it is an optimal detection statistics testing a statistical hypothesis\cite{2499}, \cite{lrr-2009-2}. The key idea of the filtering is to calculate the correlation between the output signal and a modeled GW waveform. We suppose that the output data from each of the GW telescopes is
\begin{equation}\label{ken-thesis-eq:26}
 s^{X}(t) = h^{X}(t)+n^{X}(t),
\end{equation}
where
$n^{X}(t)$ is the noise of the X-th GW telescope. The noise of a GW telescope is assumed to be stationary and Gaussian. It is characterized by the noise power spectral density(PSD)
$S^{X}_{n}(f)$ defined by
\begin{equation}\label{ken-thesis-eq:27}
  \left\langle \tilde{n}^{X}(f)\tilde{n}^{Y*}(f')\right\rangle\equiv \delta^{XY}\delta(f-f')S^{X}_{h}(f).
\end{equation}

A likelihood ratio, the ratio of the hypothesis that GW signal 
$h$ contains in the output of a GW telescope to null hypothesis, provides the output of the matched filtering. The likelihood ratio is defined by
\begin{equation}\label{ken-thesis-eq:29}
 \Lambda(h)=\frac{P(s|h)}{P(s|0)}=\frac{e^{-(s^{X}-h^{X}|s^{X}-h^{X})/2}}{e^{-(s^{X}|s^{X})}},
\end{equation}
where the inner product of the data
$a(t)$ and
$b(t)$ is defined by
\begin{equation}\label{ken-thesis-eq:28}
  \left(a(t)|b(t)\right)\equiv 4\mathrm{Re}\int^{\infty}_{0}\frac{\tilde{a}(f)\tilde{b}^{*}(f)}{S_{n}(f)}df.
\end{equation}
A log-likelihood ratio, the logarithmic form of the likelihood ratio, is expressed by
\begin{equation}\label{ken-thesis-eq:30}
 \mathrm{ln}\Lambda=(s|h)-\frac{1}{2}(h|h).
\end{equation}
The estimation of the GW parameters are evaluated by maximizing Eq.(\ref{ken-thesis-eq:30}) over GW parameters.

A coherent search method employs the value summed up with each of the log-likelihood ratios calculated by the use of the output of corresponding GW telescopes\cite{PhysRevD.83.084002}. The log-likelihood ratio for the multiple telescopes is given by
\begin{equation}\label{ken-thesis-eq:32}
 \mathrm{ln}\Lambda=A^{\mu}(\bm{s}|\bm{h}_{\mu})-\frac{1}{2}A^{\mu}M_{\mu\nu}A^{\nu},
\end{equation}
where the matrix
$M_{\mu\nu}$ is defined by
\begin{equation}\label{ken-thesis-eq:33}
 M_{\mu\nu}\equiv(\bm{h}_{\mu}|\bm{h}_{\nu}).
\end{equation}
The inner product of the multiple data is expressed by the sum of each of the inner products:
\begin{equation}\label{ken-thesis-eq:31}
 \left(\bm{a}|\bm{b}\right)\equiv \sum_{X}\left(a^{X}|b^{X}\right).
\end{equation}

Since the oscillation of a GW in the sensitive frequency band of a GW telescope is fast enough, the phases
$h_{0}$ and $h_{\pi/2}$ is regarded as orthogonal each other. The orthogonality of phases leads to express the inner product of phases as followings:
\begin{eqnarray}
 \left(h^{X}_{0}|h^{X}_{\pi/2}\right)&=&0,\label{ken-thesis-eq:34}\\
 \left(h^{X}_{0}|h^{X}_{0}\right)&=&\left(h^{X}_{\pi/2}|h^{X}_{\pi/2}\right)\equiv (\sigma^{X})^{2}.\label{ken-thesis-eq:35}
\end{eqnarray}
Therefore, Eq.(\ref{ken-thesis-eq:33}) can be expressed in terms of the inner product of phases and beam-pattern functions:
\begin{equation}\label{ken-thesis-eq:36}
 M_{\mu\nu}=
  \begin{pmatrix}
  A & C & 0& 0 \\
   C & B &0 &0\\
   0&0&A&C\\
   0&0&C&B
  \end{pmatrix},
\end{equation}
where
\begin{eqnarray}
 A&=&\sum_{X}(\sigma^{X}F^{X}_{+})^{2},\label{ken-thesis-eq:37}\\
 B&=&\sum_{X}(\sigma^{X}F^{X}_{\times})^{2},\label{ken-thesis-eq:38}\\
 C&=&\sum_{X}(\sigma^{X}F^{X}_{+})(\sigma^{X}F^{X}_{\times}).\label{ken-thesis-eq:39}
\end{eqnarray}
According to Eq.(\ref{ken-thesis-eq:36}), the log-likelihood ratio Eq.(\ref{ken-thesis-eq:32}) can be decomposed into two parts;
\begin{eqnarray}
 \mathrm{ln}\Lambda_{1}&=&A^{i}(\bm{s}|\bm{h}_{i})-\frac{1}{2}A^{i}M_{ij}A^{j},\label{ken-thesis-eq:41}\\
 \mathrm{ln}\Lambda_{2}&=&A^{k}(\bm{s}|\bm{h}_{k})-\frac{1}{2}A^{k}M_{kl}A^{l},\label{ken-thesis-eq:42}
\end{eqnarray}
where
$i,j=0,1$ and
$k,l=2,3$.

\section{regularization of a inverse problem}\label{ken-short_thesis-sec:regul-inverse-probl}
The parameter estimation requires to solve a inverse problem with a maximum likelihood method. The estimation of a GW amplitude is obtained by maximizing Eq.(\ref{ken-thesis-eq:32}) over parameter space of a GW:
\begin{equation}\label{ken-thesis-eq:40}
 \hat{A}^{\mu} = M^{\mu\nu}\left(\bm{s}|\bm{h}_{\nu}\right),
\end{equation}
where 
$\hat{A}^{\mu}$ are the estimated parameters of a GW amplitude and
$M^{\mu\nu}$ is the inverse of the inverse operator
$M_{\mu\nu}$. The solution of the inverse problem, however, often deteriorates the accuracy of a parameter space because of the degeneracy of the parameters to estimate. The degeneracy, called ill-posed problem\cite{0264-9381-23-19-S05}, is caused by rank deficiency of the inverse operator. In the literature of the estimation of GW parameters, the rank deficiency of the inverse operator happens when the array of the
beam-pattern functions of
$+$ mode
${\bf F}_{+}\equiv[F^{1}_{+}, \cdots, F^{n}_{+}]$ is approximately in proportion to these of the
$\times$ mode
${\bf F}_{\times}\equiv[F^{1}_{\times}, \cdots, F^{n}_{\times}]$, where
$n$ is the number of GW telescopes. As a result, similar response functions is detected over GW telescopes and the GW modes are degenerated each other. Since the degeneracy of GW modes amplifies the noise, called amplified noise, the accuracy of the parameters of GW amplitude deteriorates. To avoid the the ill-posed problem, a regularization method\cite{engl1996regularization}, \cite{ENGL1988395}, \cite{10.2307/2008325} provides the effective solution to recover the rank deficiency of the inverse operator by applying a correction term to a likelihood. For a GW analysis, M. Rakhmanov has indicated the possibility that the reconstruction of the GW can be improved by applying Tikhonov regularization method\cite{0264-9381-23-19-S05}. His work, however, couldn't optimize full of the parameters introduced by the regularization method, named regulator parameters. To optimize all of the regulator parameters and apply for the modeled GW, we develop the new regularization method by using two methods; the method of Lagrange multiplier with Karush-Kuhn-Tucker (KKT) conditions\cite{Boyd:2004:CO:993483} and the novel parameter choice method to obtain the full optimization of the regulator parameters.

In this section, we discuss the method to optimize full of regulator parameters to suppress the amplified noise caused by a ill-posed problem. In Sec.\ref{ken-short_thesis-sec:regul-inverse-probl}, we explain the regularization method using the Eq.(\ref{ken-thesis-eq:41})

\subsection{maximizing the likelihood with a regularization method}\label{ken-short_thesis-sec:maxim-likel-with}
To suppress the amplified noise, the key idea is to add a correction term to the log-likelihood ratio:
\begin{equation}\label{ken-thesis-eq:43}
  \mathrm{ln}\Lambda_{1g}=A^{i}(\bm{s}|\bm{h}_{i})-\frac{1}{2}A^{i}M_{ij}A^{j} - \|A^{i}\Omega_{ij}(\omega[k])A^{j}\|,
\end{equation}
where the regulator
$\Omega_{ij}$ is a
$2\times 2$ matrix of regulator parameters
$\omega[k]$,
the
$k$ is the number of regulator parameters. Index
$g$ indicates the regularized log-likelihood ratio. The estimation
$\hat{A}^{i}_{g}$, which are the parameters of a GW amplitude with regularization method, is evaluated from maximizing Eq.(\ref{ken-thesis-eq:43}) over parameters of a GW amplitude
\begin{equation}\label{ken-thesis-eq:45}
 \hat{A}_{g}^{i} = M_{g}^{ij}(\bm{s}|\bm{h}_{i}),
\end{equation}
where
\begin{equation}\label{ken-thesis-eq:46}
 M_{g,ij}\equiv M_{ij} + 2\Omega_{ij}(\omega[k]).
\end{equation}

Eq.(\ref{ken-thesis-eq:46}) clearly shows the inverse operator
$M_{ij}$ is corrected by the regulator and prevents the inverse operator from reducing the number of rank. The numerical expression of matched filtering of output can be decomposed into a GW signal and the noise of GW telescopes:
\begin{equation}\label{ken-thesis-eq:47}
 (\bm{s}|\bm{h}_{i}) = A^{j}M_{ij} + (\bm{n}|\bm{h}_{i}).
\end{equation}

The distinct decomposition into a GW signal and a noise couldn't be calculated because the GW signal in output of GW telescopes is never known. Note that the noise of the matched filtering
$(\bm{n}|\bm{h}_{i})$ is calculated by the variance of the values of the matched filtering under the Bayesian inference. To make it clear, we write the mismatching as
$(\bm{n}|\bm{h}_{i}) \equiv \Delta(\bm{s}|\bm{h}_{i})$.

The numerical expression of the residual noise of a GW amplitude is obtained as follows:
\begin{widetext}
 
\begin{eqnarray}
 \left\|\hat{A}_{g}^{i} - A^{i}\right\|^{2} &=& 4\left\|M_{g}^{ij}\Omega_{ji'}(\omega([k]))A^{i'}\right\|^{2} + \sum_{i,i'}M_{g}^{ij}M_{g}^{i'j'}\left\langle\Delta(\bm{s}|\bm{h}_{j})\Delta(\bm{s}|\bm{h}_{j'})\right\rangle\label{ken-report-eq:85}\\
 &\equiv& B(\omega[k], A^{i}) + N(\omega[k], \Delta(\bm{s}|\bm{h}_{i}))\label{ken-short_thesis-eq:3},
\end{eqnarray}
\end{widetext}
where
$i',j'=0,1$,
$\langle\rangle$
is the ensemble average of the number of iterations performed in the procedure of the Bayesian inference, and
$\|\cdot^{i}\|^{2}\equiv \sum_{i}\left\langle(\cdot^{i})^{2}\right\rangle$. The first term
$B(\omega[k], A^{i})$
corresponds to the bias noise introduced by the regulator and the second term
$N(\omega[k], \Delta(\bm{s}|\bm{h}_{j}))$
corresponds to the amplified noise suppressed by the use of a regularization method, called reduced noise. With an appropriate choice of regulator parameters, the sum of the two noise is possible to be smaller than the amplified noise even if the bias noise, which behaves additional noise of a estimation, is introduced.

To find the optimized regulator parameters, we must consider the method to minimize the residual noise Eq.(\ref{ken-report-eq:85}). The minimization is archived by finding the local minimum of the residual noise over the regulator parameter space. One should be paid attention; In the case that the amount of the bias noise is larger than the reduced noise as a result of the optimization of regulator parameters, the intrinsic parameters of a GW lies outside of the credible regions of the probability distribution. It indicates that an appropriate restriction for determining the regulator parameters is required. We propose the requirement as a following:
\begin{equation}\label{ken-thesis-eq:51}
 \frac{B(\omega[k], A^{i})}{N(\omega[k], \Delta(\bm{s}|\bm{h}_{j}))} \leq C,
\end{equation}
where
$C$ is the real and positive user-defined value that is less than 1.

The minimization of the residual noise under the condition of Eq.(\ref{ken-thesis-eq:51}) is performed by the method of Lagrange multiplier with Karush-Kuhn-Tucker (KKT) conditions\cite{Boyd:2004:CO:993483}:
\begin{widetext}
 \begin{eqnarray}\label{ken-thesis-eq:44}
 \frac{\partial}{\partial \omega[k]}\left\|\hat{A}_{g}^{i} - A^{i}\right\|^{2} + \lambda \frac{\partial}{\partial \omega[k]}\left(B(\omega[k], A^{i}) - CN(\omega[k], \Delta(\bm{s}|\bm{h}_{j})\right)&=&0\ \ \mathrm{for}\ \mathrm{all}\ k,\\
  \frac{B(\omega[k], A^{i})}{N(\omega[k], \Delta(\bm{s}|\bm{h}_{j}))} &\leq& C,\nonumber
\end{eqnarray}
\end{widetext}
where the real positive number
$\lambda$
is a Lagrange multiplier.

While Eq.(\ref{ken-thesis-eq:44}) has a lot of solutions in regulator parameters, the optimized regulator parameters are able to be determined uniquely by the use of following criteria: (i) The complex values of the regulator parameters must be excluded. (ii) the regulator parameters in which Eq.(\ref{ken-thesis-eq:44}) indicates 0 or infinity by substituting must be excluded. (iii) the regulator parameters are chosen so that the real value of the Lagrange multiplier is the highest one satisfying the criteria (i) and (ii).

\subsection{a-posteriori parameter choice rule}\label{ken-thesis-sec:impr-post-param}
The evaluation of Eq.(\ref{ken-thesis-eq:44}) is essentially impossible in the Bayesian inference because the method to determine optimized regulator parameters includes a fatal defect; the intrinsic parameters of a GW amplitude, which needs to be known for evaluating the bias noise of Eq.(\ref{ken-report-eq:85}), are never known a-priori. The key idea to avoid this problem is to replace the intrinsic parameters of a GW amplitude  in the residual noise with the estimated parameters:
\begin{equation}\label{ken-thesis-eq:55}
  A^{i}\rightarrow M^{ij}\left(\bm{s}|\bm{h}_{j}\right).
\end{equation}
This method is called a-posteriori parameter choice rule\cite{engl1996regularization}. By adapting the parameter choice rule, the optimized regulator parameters can be calculated without using intrinsic parameters of a GW amplitude. The estimated parameters are obtained in each of the iterations of the non-regularized Bayesian inference. The proof of the idea is shown in \cite{engl1996regularization}\cite{ENGL1988395} in detail, but in the case of the Eq.(\ref{ken-thesis-eq:44}), the essence of the validity of the parameter choice rule can be explained as follows; the ensemble average of the residual noise calculated by the intrinsic parameters are essentially equivalent to the one calculated by estimated parameters because the difference between the intrinsic parameters and estimated parameters is canceled out over the iterative calculation of the Bayesian inference. As a result, the optimized regulator parameters are obtained by solving following equations:
\begin{widetext}
 \begin{eqnarray}\label{ken-thesis-eq:52}
 \frac{\partial}{\partial \omega[k]}\left\|\hat{A}_{g}^{i} - M^{ij}\left(\bm{s}|\bm{h}_{j}\right)\right\|^{2} + \lambda \frac{\partial}{\partial \omega[k]}\left(B(\omega[k], M^{ij}\left(\bm{s}|\bm{h}_{j}\right)) - CN(\omega[k], \Delta(\bm{s}|\bm{h}_{i})\right)&=&0\ \ \mathrm{for}\ \mathrm{all}\ k,\\
  \frac{B(\omega[k], M^{ij}\left(\bm{s}|\bm{h}_{j}\right))}{N(\omega[k], \Delta(\bm{s}|\bm{h}_{i}))} &\leq& C.\nonumber
\end{eqnarray}
\end{widetext}

\subsection{regulator}\label{ken-thesis-sec:formalism-regulator}
In this section, we will give the regulator matrix. As mentioned before, the rank deficiency of an inverse operator happens when one of the proper vectors of the inverse operator is approximately vanished. To compensates the lack of the proper vector, two kinds of regulator matrices are considered in this thesis. We call the formar regulator as type 1 and the latter one as type 2.

(i) Tikhonov type regulator (type 1):
\begin{equation}\label{ken-short_thesis-eq:1}
 \Omega = \begin{pmatrix}
           w & 0 \\
           0 & w 
          \end{pmatrix},
\end{equation}
where
$w$ is the regulator parameter which has real value. This is a traditional regulator matrix form being used for a lot of ill-posed problem.

(ii) Symmetric trace-free type regulator (type 2):
\begin{equation}\label{ken-short_thesis-eq:2}
 \Omega = \begin{pmatrix}
           w & we \\
           we & we^{2} 
          \end{pmatrix},
\end{equation}
where
$w$
and
$e$ are the regulator parameters which have real values. This type of the regulator is introduced for the first time. This regulator is based on the following two ideas; the rank of the regulator is at most 1 because the number of lack of the proper vector is at most 1, and the regulator matrix has symmetry which corresponds to the symmetric matrix of the inverse operator.

\section{analysis}\label{ken-short_thesis-sec:analysis}
This section first presents an overview of our regularized data analysis and the Bayesian inference algorithm. We also discuss the states of a software-injected signal of a GW from CBC; then the results of our regularized data analysis are presented and discussed.

\subsection{Data analysis}\label{ken-short_thesis-sec:nest-sampl-algor}
Bayesian inference for a software-injected GW signal is performed using the nested sampling algorithm, which has been described by Skilling\cite{skilling2006nested} as a reversal of typical Bayesian inference such as Markov chain Monte Carlo methods\cite{gelman2013bayesian}. Note that a detailed study into the parameters estimation of a GW using the nested sampling algorithm has been performed previously\cite{PhysRevD.81.062003}. MultiNest implements Bayesian inference based on the nested sampling algorithm\cite{2009MNRAS.398.1601F}\cite{2013arXiv1306.2144F} and resolves several problems using the nested sampling algorithm. For example, the ellipsoid sampling method provides more efficient implementation of the Bayesian inference by approximating the iso-likelihood contour of a point to be replaced by a D-dimensional (D is the number of search parameters) ellipsoid determined by the covariance matrix of the current set of the live points. Furthermore, the recursive clustering method allows us to evaluate the multimodal posterior probability distribution by dividing an approximated ellipsoid into multiple approximated ellipsoids using K-means algorithm\cite{Shaw11072007}\cite{2008MNRAS.384..449F}. In addition to MultiNest, PyMultiNest software\cite{refId0}, a Python interface for MultiNest, is also used in the data analysis. Note that MultiNest users must specify a few proper parameters, i.e., the number of live points and the tolerance, and set the prior and log-likelihood functions. In this study, the number of live points is 1000 for all data analyses and the tolerance is set to 0.0001 for the data analysis without a regulator (non-regularized data analysis) and 0.05 for the data analysis with regulator-optimized analysis (regularized data analysis).

We must evaluate Eq.(\ref{ken-thesis-eq:52}) for the regularized data analysis to determine optimized regulator parameters. However, the evaluation of Eq.(\ref{ken-thesis-eq:52}) requires another algorithm in addition to the nested sampling algorithm. As discussed previously, the degree of mismatch of the matched filtering evaluated relative to the accuracy of the phase parameters, such as a chirp mass and a coalescence time when a GW passes though the center of the earth, is requires to determine optimized regulator parameters. These mismatch values are obtained by evaluating the accuracy of the phase parameters obtained by the non-regularized data analysis. The regularized data analysis  is performed as follows. First, non-regularized data analysis is performed to obtain the covariance matrix of the phase parameters. Here, we also obtain estimated amplitude parameter values and corresponding credible regions to confirm the accuracy improvement gained by the amplitude parameters obtained in the regularized data analysis. Second, we obtain the distribution of
$h_{\mu}$ by substituting the distribution of phase parameters generated by multivariate random distribution of the covariance matrix of the phase parameters into the matched filtering. Here, the number of samples in the distribution set of
$h_{\mu}$ are 10000 in this thesis. Third, the variance of the
$h_{\mu}$ is calculated using the distribution set.

The details the data analysis are as follows: the number of search parameters of the GW is six, inclination $\iota$, luminosity distance $r$, polarization $\Phi$, initial phase $\phi_{\mathrm{coal}}$, chirp mass $M_{c}$, and coalescence time $t_{\mathrm{coal}}$ when a GW passes through the center of the earth. To reduce data analysis time, we ignore the post-Newtonian parameters of a GW. Note that all of the prior distribution of the search parameters are flat distributions. In addition, the data analysis result is expressed in terms of physical parameters: however, the data analysis is implemented using the terms of amplitude parameters
$A^{1}-A^{4}$. We convert these amplitude parameters into physical parameters after data analysis is performed. The range of the amplitude parameters is
$A^{i}\in [-10^{-15},10^{-15}]$. The range of the inclination expressed as a cosine is
$\cos\iota \in [-1,1]$, luminosity distance is
$r\in[10:1000]$Mpc, polarization is
$\Phi\in[0,\pi]$ and initial phase is
$\phi_{\mathrm{coal}}\in[0,\pi]$. To reduce data analysis time, the range of a chirp mass is set to
$\pm 1 M_{solar}$ of the input value, and the coalescence time range is set to
$\pm 0.1 s$ of the input value. These range settings are appropriate because the regularization method helps in reducing the accuracy of amplitude parameters, and the accuracy of the phase parameters is not affected significantly.

Both regulators (Eq.(\ref{ken-short_thesis-eq:1}) and Eq.(\ref{ken-short_thesis-eq:2})) are applied to the regularized data analysis to compare the differences in the reduction of the accuracy of the amplitude parameters. Here, the user-defined value of
$C$ in Eq.(\ref{ken-thesis-eq:51}) is set to 0.1. To reduce the computational cost of the type 2 regulator (Eq.(\ref{ken-short_thesis-eq:2})), we exclude the regulator parameter
$e$ to evaluate the requirement in Eq.(\ref{ken-thesis-eq:51}). The validity of the exclusion of the calculation can be explained by the fact that the variable
$e$ in the type 2 regulator does not have strong dependency on the strength of the regulator. The strength of the regulator primarily affects the amount of Eq.(\ref{ken-thesis-eq:51}).

\subsection{Software-injected data}\label{server-thesis-sec:cond-impl}
The software-injected GW signal is examined herein.  In this study, the data analysis is implemented using two types of GW signal. The different status of these signals reflects the amplitude parameters because we are interested in the difference in the reduction of the amplified noise relative to the amplitude parameters of a GW. In the following sections, we first describe the common states of the software-injected GW signal; then, we discuss the difference in the amplitude parameters.
\subsubsection{Common states}\label{ken-short_thesis-sec:common-status}
We consider that the network of GW telescopes comprises the LIGO Livingston and the LIGO Hanford\cite{0264-9381-32-7-074001}. The location of the GW telescopes is the actual geographic state\cite{1996gr.qc.....7075A} and the noise power spectrum of the GW telescopes uses the theoretical function of aLIGO\cite{lrr-2009-2}. The properties of the noise of GW telescopes are assumed be both stationary and Gaussian, and the integrated SNR of the software-injected GW signal is set to
$20\pm 0.5$. The common states of the software-injected GW signal are listed in Table.\ref{server-thesis-tab:2}. The details of these parameters are explained in the caption.

\begin{table}[htb]
  \begin{center}
    \begin{tabular}{|l||r|r|} \hline
     Parameters & Values &Unit \\ \hline \hline
     Observation UTC & 2006/11/01  & yyyy/mm/dd\\
      & 15:45:12 & hh:mm:ss \\
     chirp mass& 7 & solar mass \\
     coalescence time& 8& second\\
     integrated SNR & 20$\pm$0.5 & \\
     time duration& 10 &sec \\
     sampling frequency& 2048 &Hz\\
     upper frequency& 1024 &Hz\\
     cutoff frequency & 20 & Hz\\\hline
    \end{tabular}
  \end{center}
    \caption{Common state of the parameters of the software-injected GW signal. Observation UTC is the time at which the GW passes through in the Earth's center of gravity. Chirp mass and coalescence time indicate the phase parameters of the GW. Integrated SNR is the SNR in the GW signal detected by the network of GW telescopes (LIGO Hanford and LIGO Livingston.) simultaneously Time duration is the length of time of the output data of the GW telescopes. Sampling, upper and cutoff frequencies indicate the configuration of the calibrated output data.\label{server-thesis-tab:2}}
\end{table}

\subsubsection{Different states}\label{ken-short_thesis-sec:different-status}
\begin{figure}[tb]
  \centering
   \includegraphics[width=8cm]{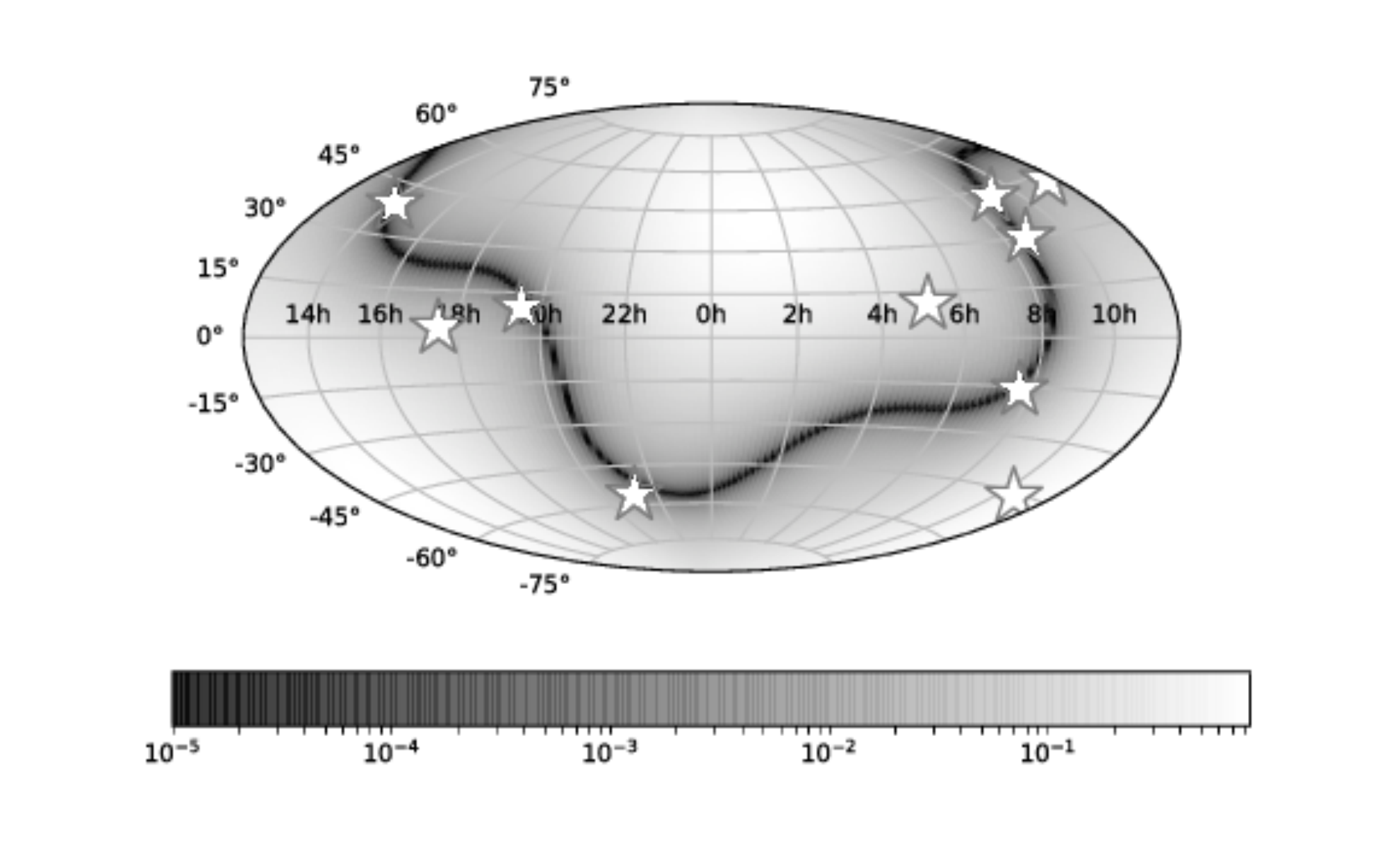}
   \includegraphics[width=8cm]{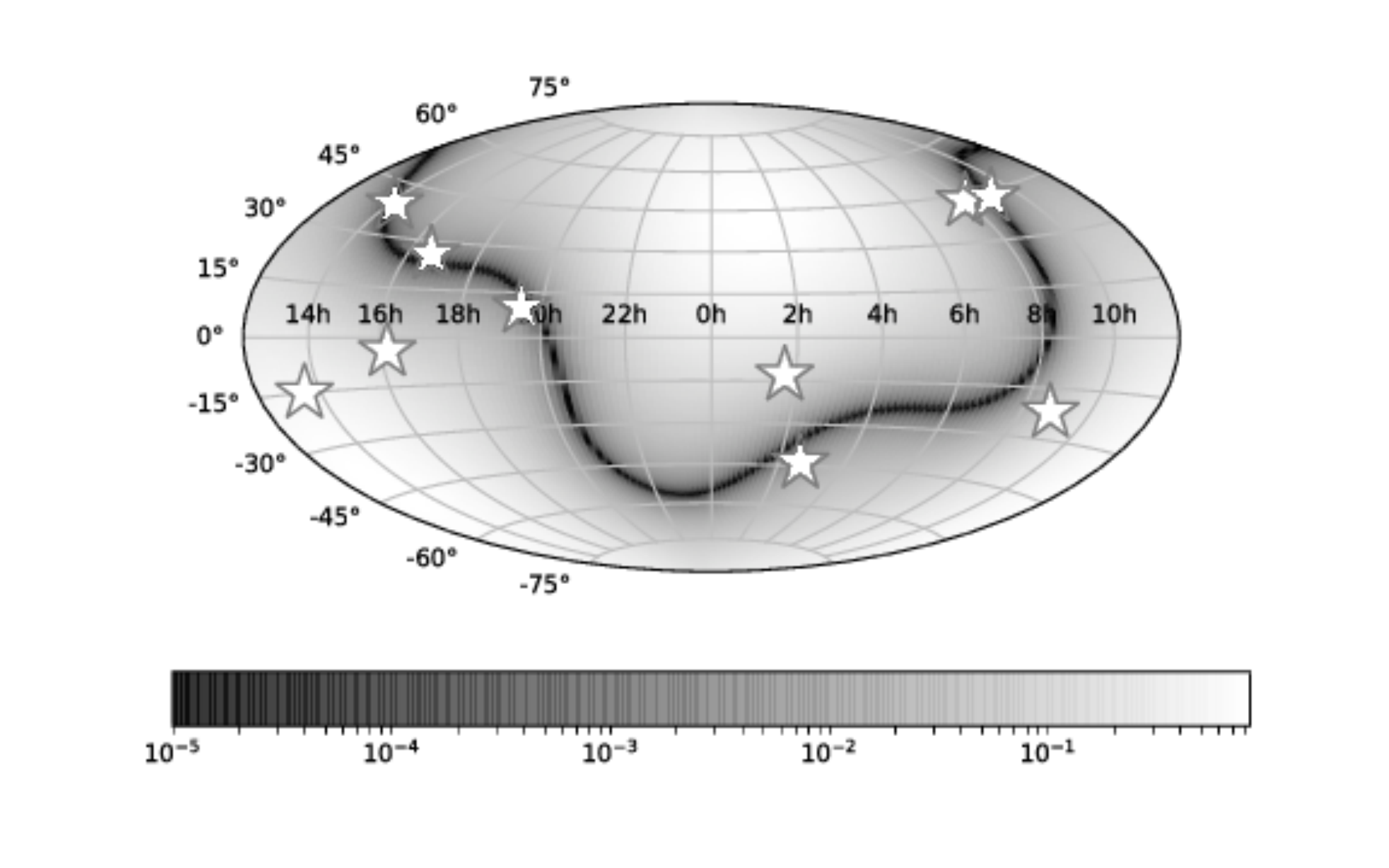}
   \caption{The sky location of the software-injected data and the determinant values of the antenna-beam pattern function. The degree of shade in both figures indicates the amount of determinant of beam-pattern function all over the sky region at the UTC specified in Table.\ref{server-thesis-tab:2}. The stars indicates the location of the target CBC of the software-injected GW signal. The first panel shows the state 1 of software-injected data and the second panels shows the same of state 2.\label{ken-short_thesis-fig:1}}
\end{figure}

Two different states are involved in the amplitude parameters of the GW signal. We are interested in reducing each amplitude parameter because the accuracy of the amplitude parameters of a GW is reduced by the regularized data analysis. The states of the amplitude parameters of the software-injected data are given in Table.\ref{ken-short_thesis-tab:3}. The luminosity distance is set to the value which the integrated SNR indicates
$20\pm$. The location of the target CBC in celestial coordinates is shown in Figure.~\ref{ken-short_thesis-fig:1}, which also illustrates the determinant value of the beam-pattern functions because the influence of ill-posed of the inverse operator increases relative to the reduction of the determinant value.
\begin{table}[htb]
  \begin{center}
    \begin{tabular}{|l||r||r|r|} \hline
     Parameters & State 1  & State 2 &Unit \\ \hline \hline
     inclination ($\cos\iota$) & 0.400 & -0.750& \\
     initial phase&-0.448 &-1.023 &Radian \\
     polarization&-0.480 &0.462 &Radian \\\hline
    \end{tabular}
  \end{center}
    \caption{States of amplitude parameters of software-injected GW signal. State 1 indicates the values of the first software-injected GW signal, and state 2 indicates the values of the second signal. The inclination data are expressed by cosine function.\label{ken-short_thesis-tab:3}}
\end{table}

\subsection{Result and discussion}\label{ken-short_thesis-sec:result}
\begin{figure*}[tb]
  \centering
   \includegraphics[width=16cm]{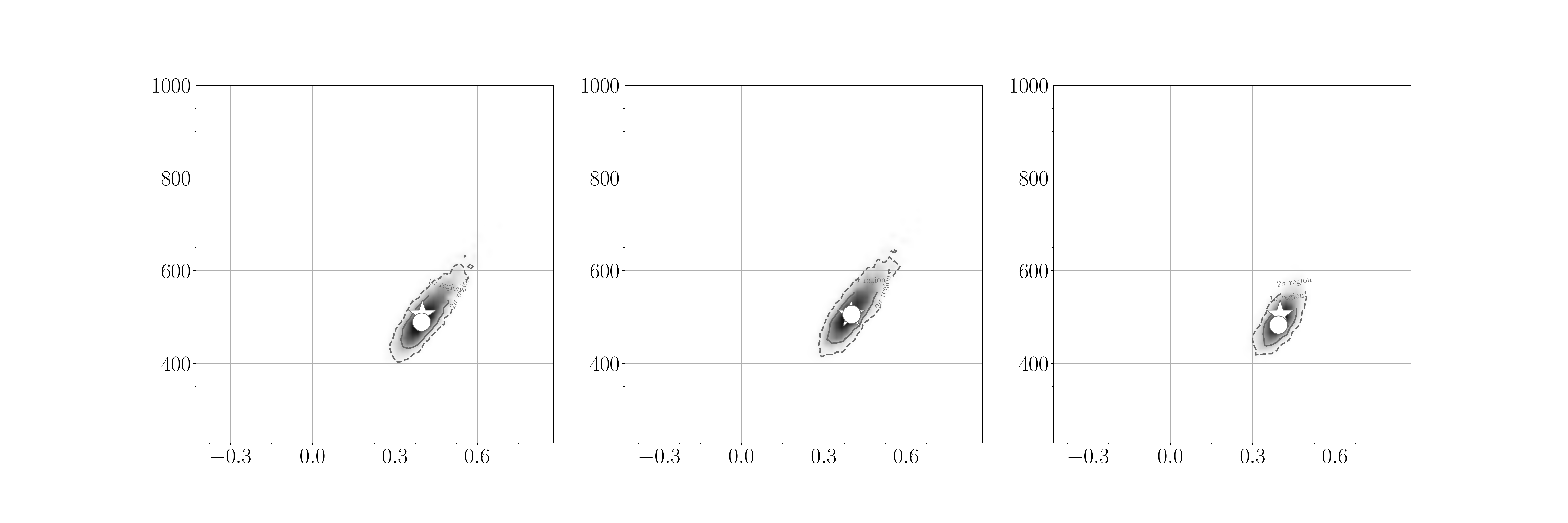}
 \includegraphics[width=16cm]{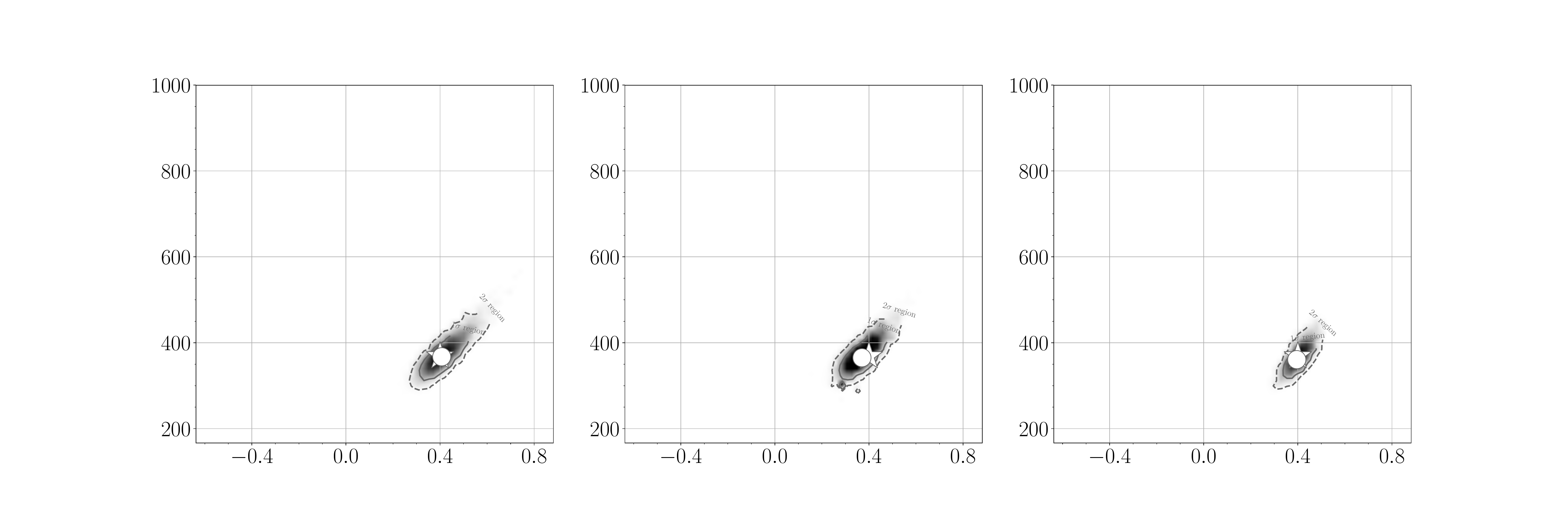}
    \includegraphics[width=16cm]{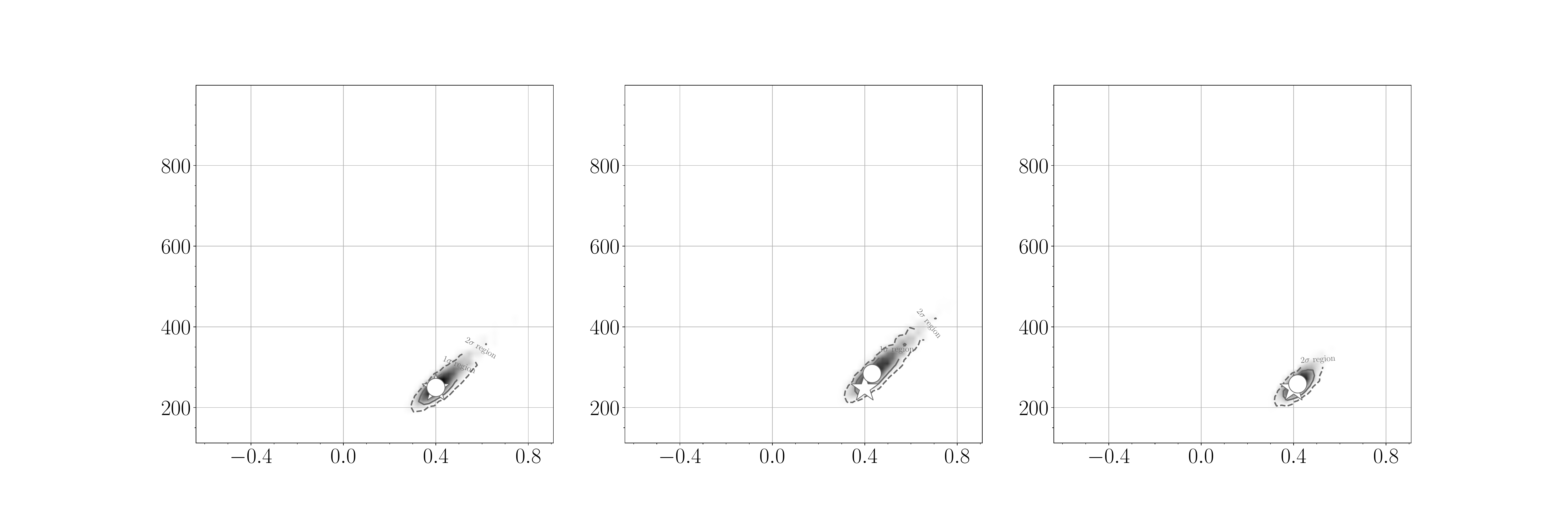}
   \caption{The credible region of the joint-probability distribution represented by profile likelihood. The state 1 of the software-injected GW signal is used. The vertical scale indicates the luminosity distance and the horizontal scale indicates the inclination expressed in cosine. The left panel of all figures shows the result of non-regularized data analysis, the center panel of all figures shows the same of regularized data analysis with the type 1 regulator and the left panel of all figures shows the same of regularized data analysis with the type 2 regulator. The star indicate the actual parameters of the software-injected GW signal and the circle indicate the maximum likelihood parameters evaluated by the data analysis respectively. The determinant values of beam-pattern function matrix are arranged $5\times10^{-1}, 1\times10^{-1}, 5\times10^{-2}$ from top to bottom.\label{ken-short_thesis-fig:8}}
\end{figure*}

\begin{figure*}[htbp]
 \centering
  \includegraphics[width=16cm]{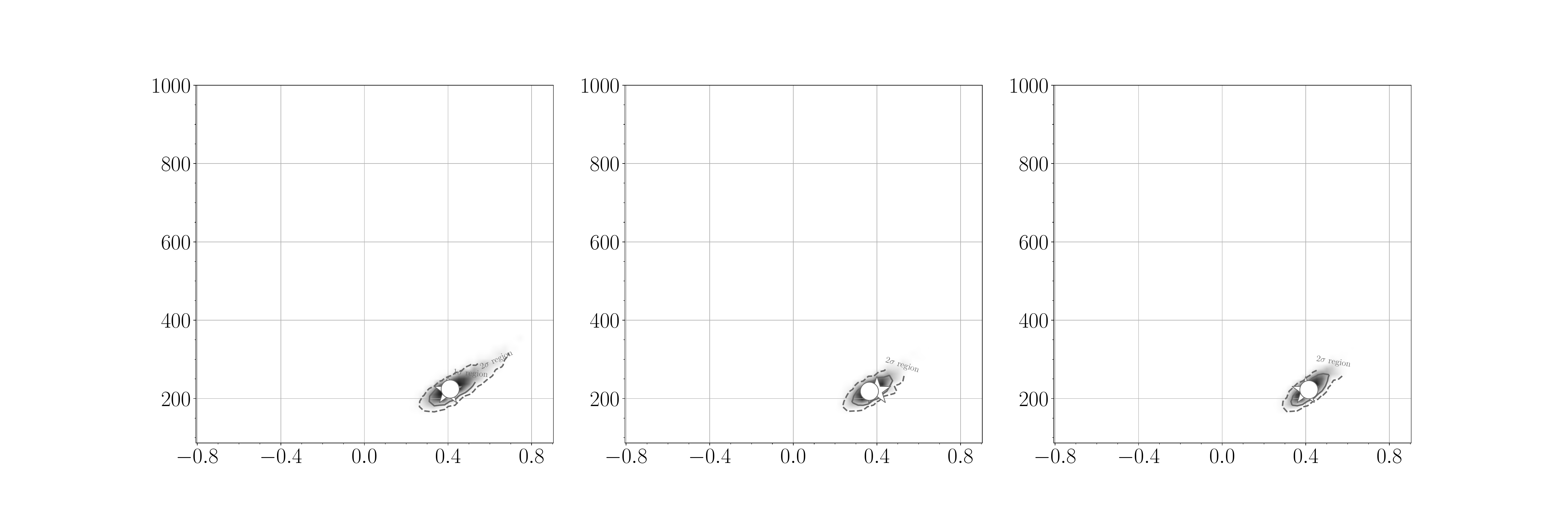}
    \includegraphics[width=16cm]{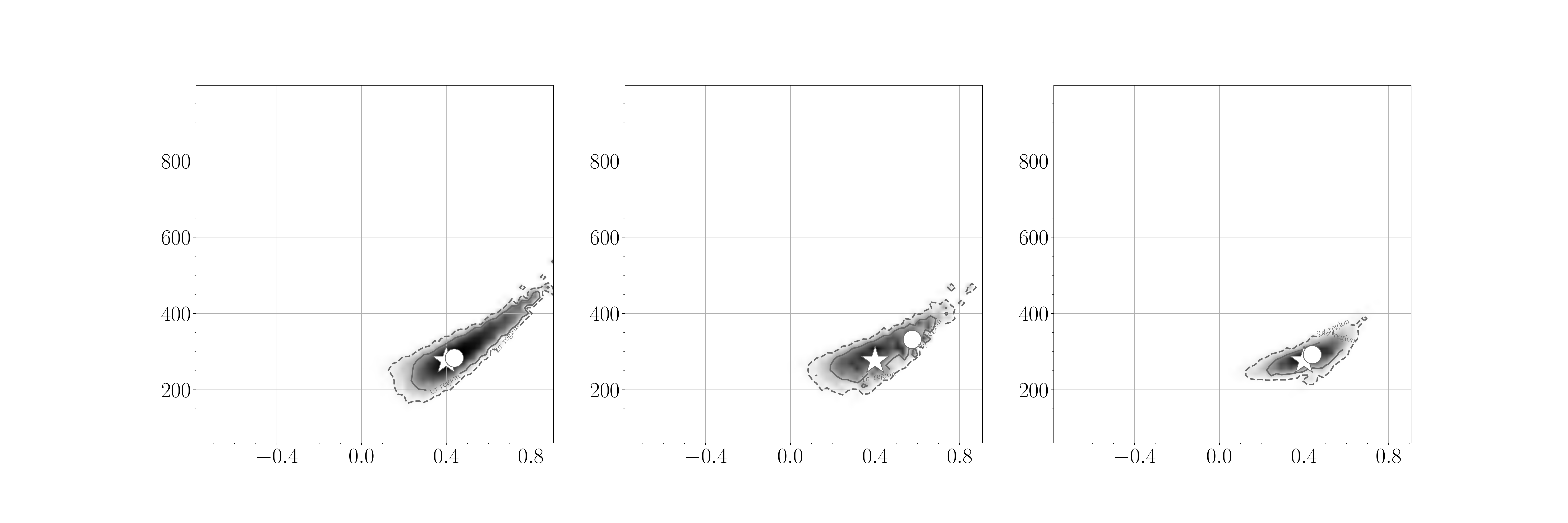}
 \includegraphics[width=16cm]{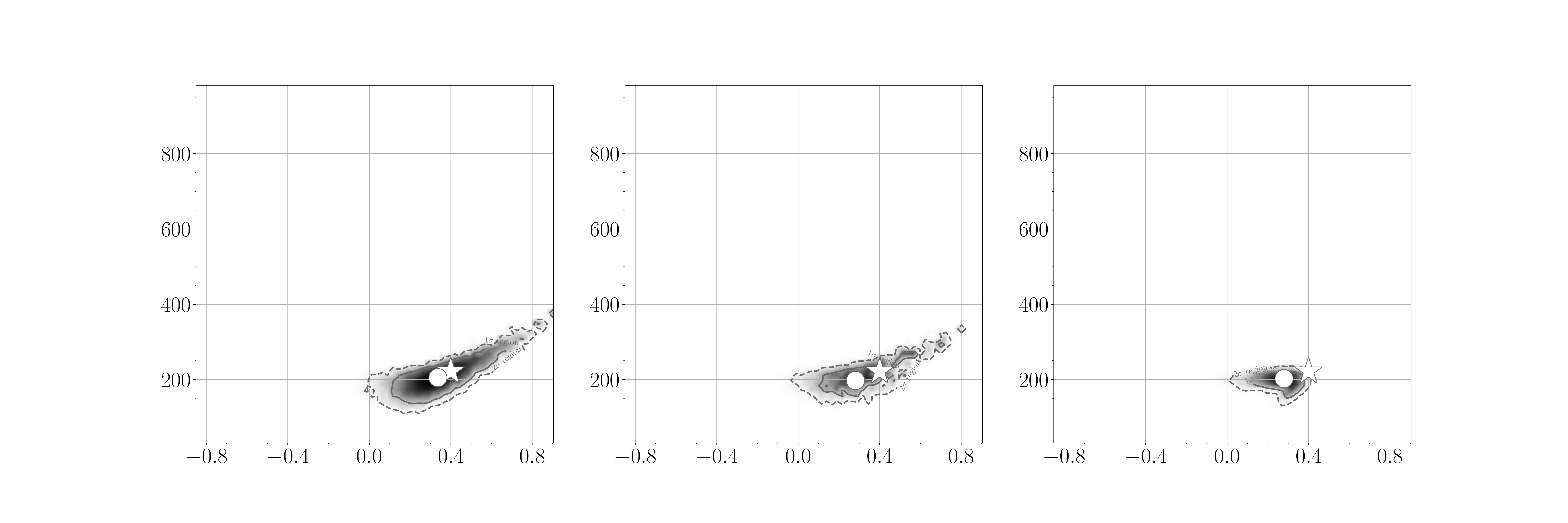}
    \caption{The probability distribution region represented by profile likelihood. The state 1 of the software-injected GW signal is used. The explanation of these figures are as same as Fig.~\ref{ken-short_thesis-fig:8}. The determinant values of beam-pattern function matrix arrange $1\times10^{-2}, 5\times10^{-3}, 1\times10^{-3}$ from top to bottom.\label{ken-short_thesis-fig:9}}
\end{figure*}

\begin{figure*}[htbp]
  \centering
    \includegraphics[width=16cm]{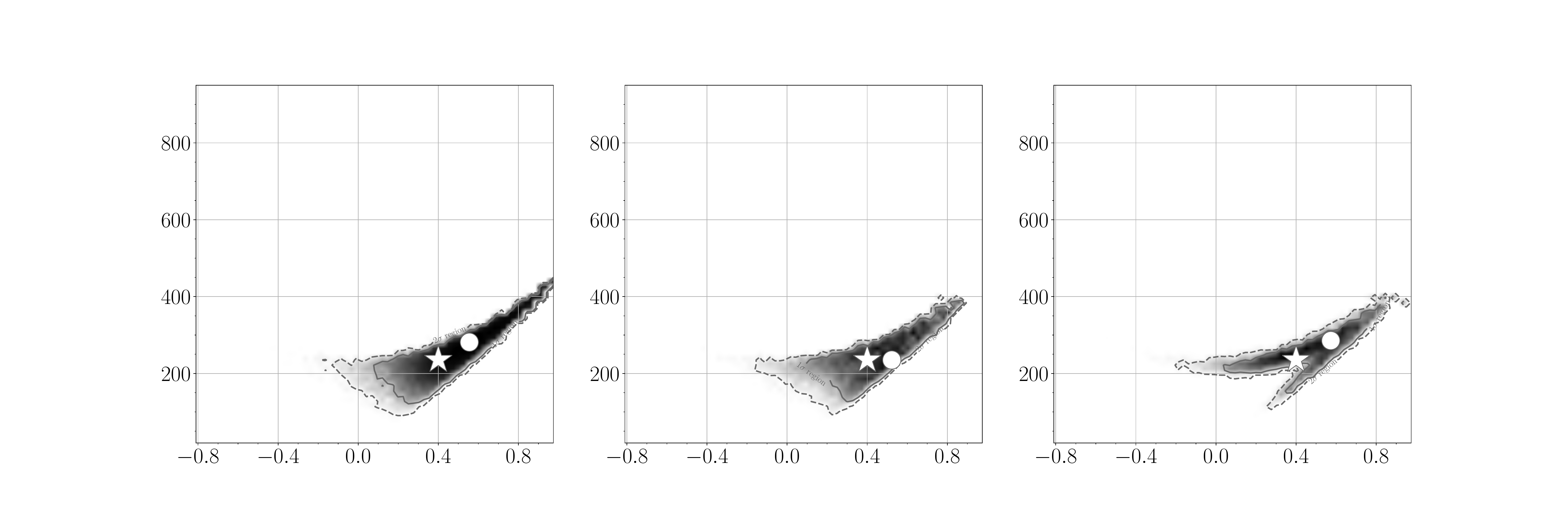}
 \includegraphics[width=16cm]{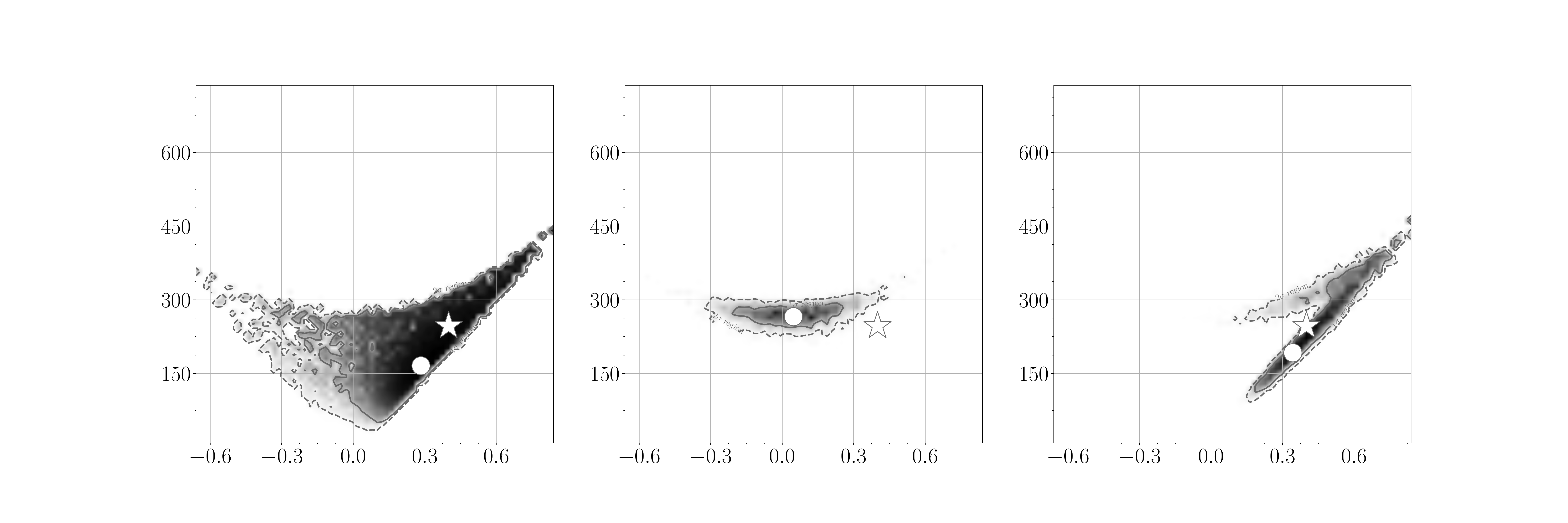}
    \includegraphics[width=16cm]{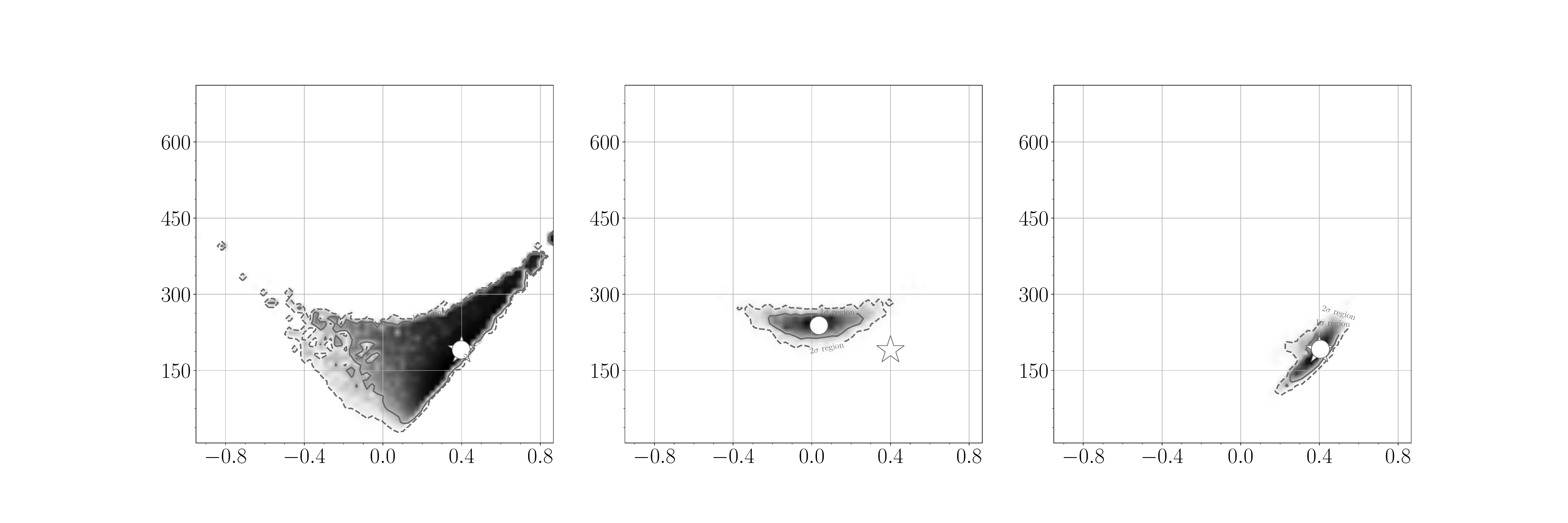}
 \includegraphics[width=16cm]{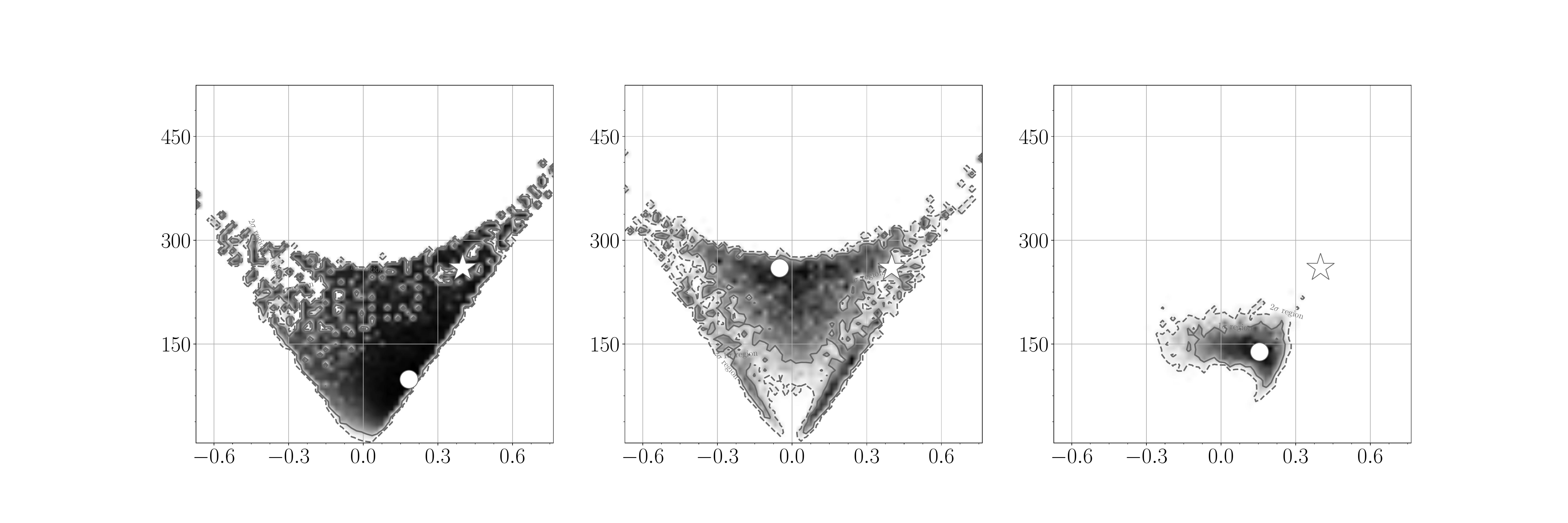}
    \caption{The probability distribution region represented by profile likelihood. The state 1 of the software-injected GW signal is used. The explanation of these figures are as same as Fig.~\ref{ken-short_thesis-fig:8}. The determinant values of beam-pattern function matrix are arranged $5\times10^{-4}, 1\times10^{-4}, 5\times10^{-5}, 1\times10^{-5}$ from top to bottom.\label{ken-short_thesis-fig:10}}
\end{figure*}

\begin{figure*}[tb]
  \centering
   \includegraphics[width=16cm]{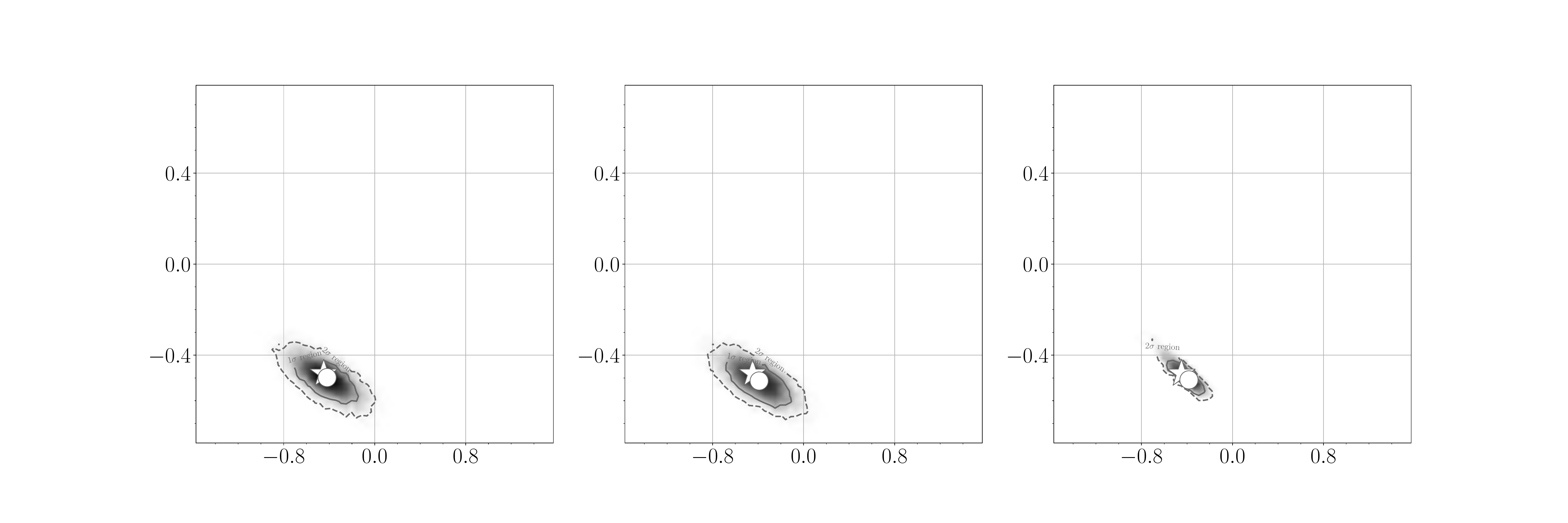}
 \includegraphics[width=16cm]{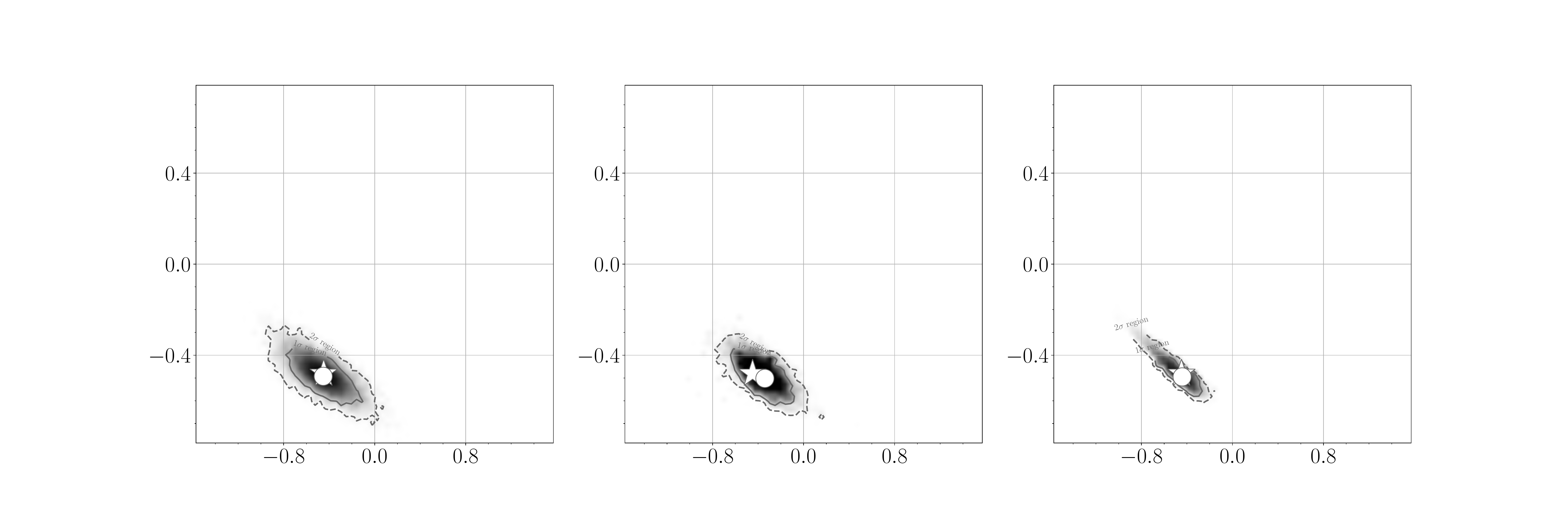}
    \includegraphics[width=16cm]{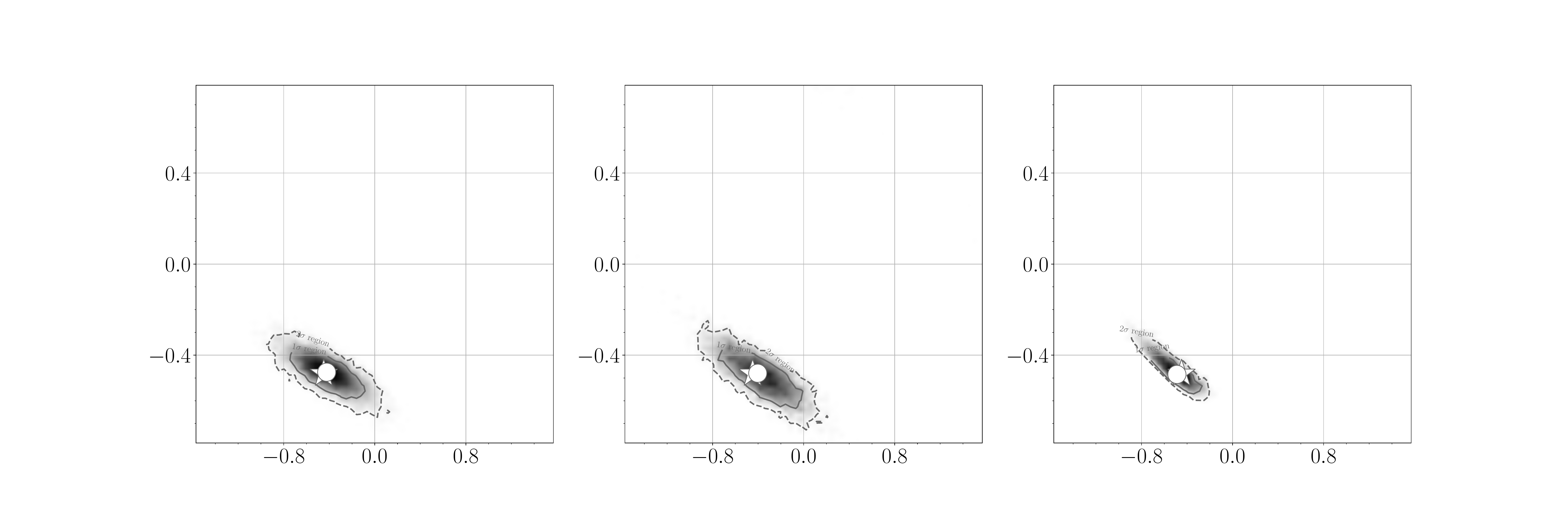}
   \caption{The probability distribution region represented by profile likelihood. The state 1 of the software-injected GW signal is used. The vertical scale the indicates initial phase and the horizontal scale indicates the polarization. The left panel of all figures shows the result of non-regularized data analysis, the center panel of all figures shows the same of regularized data analysis with the type 1 regulator and the left panel of all figures shows the same of regularized data analysis with the type 2 regulator. The star indicate the actual parameters of the software-injected GW signal and the circle indicate the maximum likelihood parameters evaluated by the data analysis respectively. The determinant values of beam-pattern function matrix are arranged $5\times10^{-1}, 1\times10^{-1}, 5\times10^{-2}$ from top to bottom.\label{ken-short_thesis-fig:11}}
\end{figure*}

\begin{figure*}[htbp]
 \centering
  \includegraphics[width=16cm]{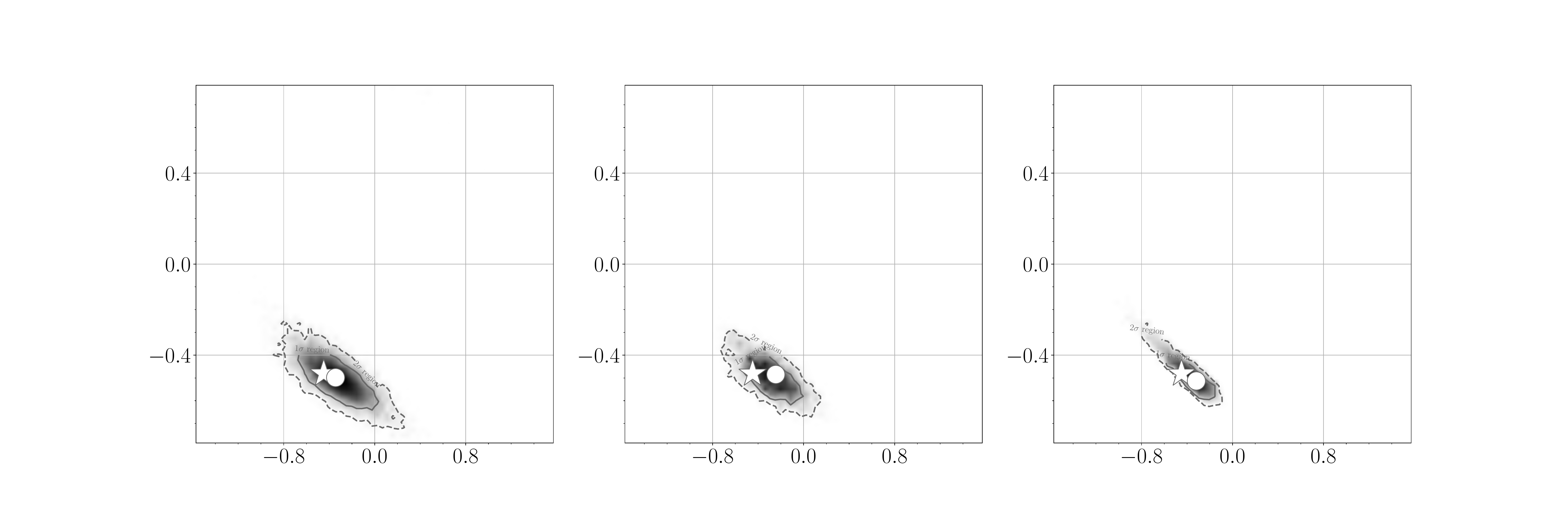}
    \includegraphics[width=16cm]{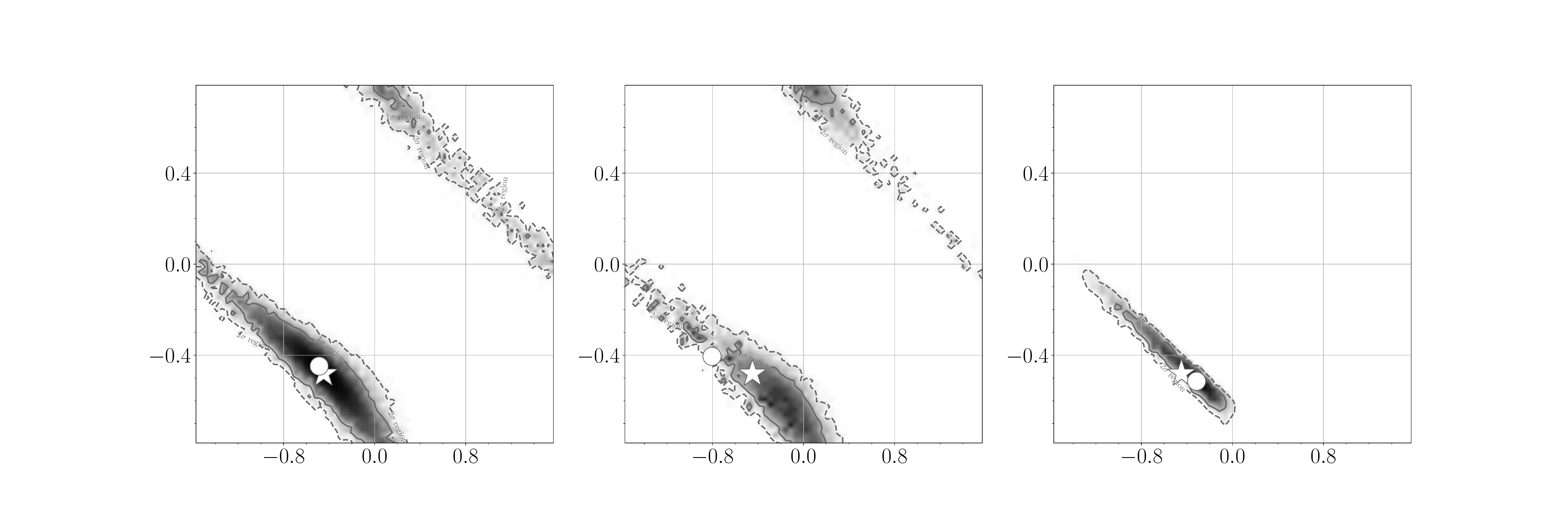}
 \includegraphics[width=16cm]{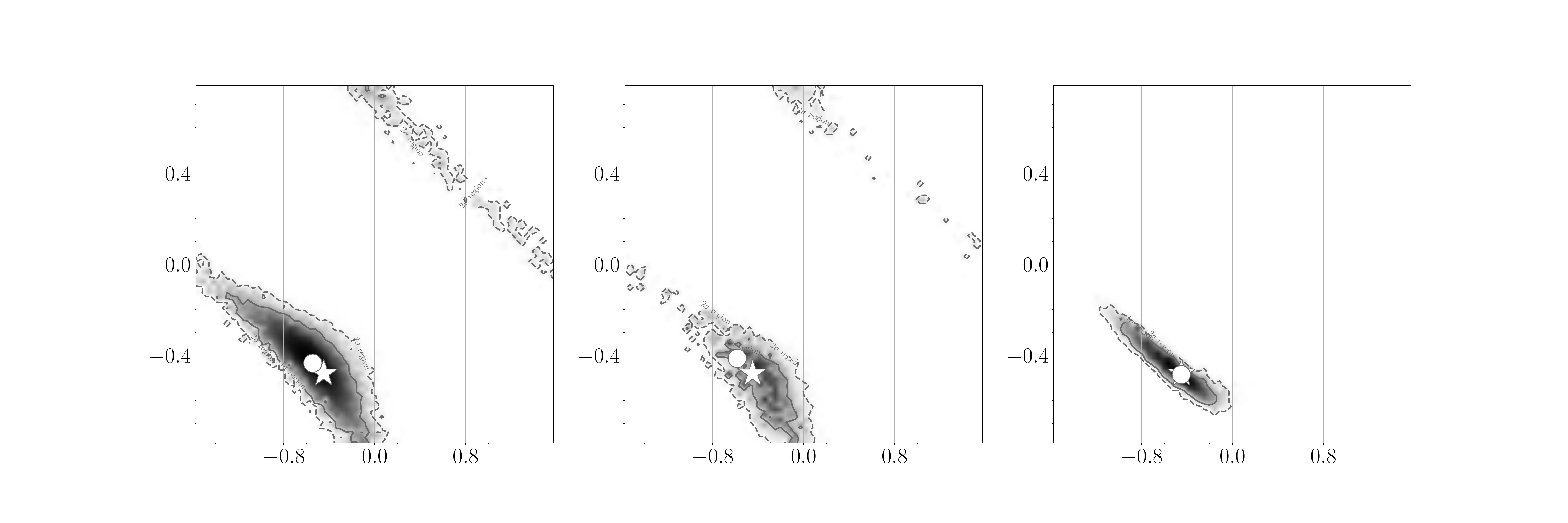}
    \caption{The probability distribution region represented by profile likelihood. The state 1 of the software-injected GW signal is used. The explanation of these figures are as same as Fig.~\ref{ken-short_thesis-fig:11}. The determinant values of beam-pattern function matrix are arranged $1\times10^{-2}, 5\times10^{-3}, 1\times10^{-3}$ from top to bottom.\label{ken-short_thesis-fig:12}}
\end{figure*}

\begin{figure*}[htbp]
  \centering
    \includegraphics[width=16cm]{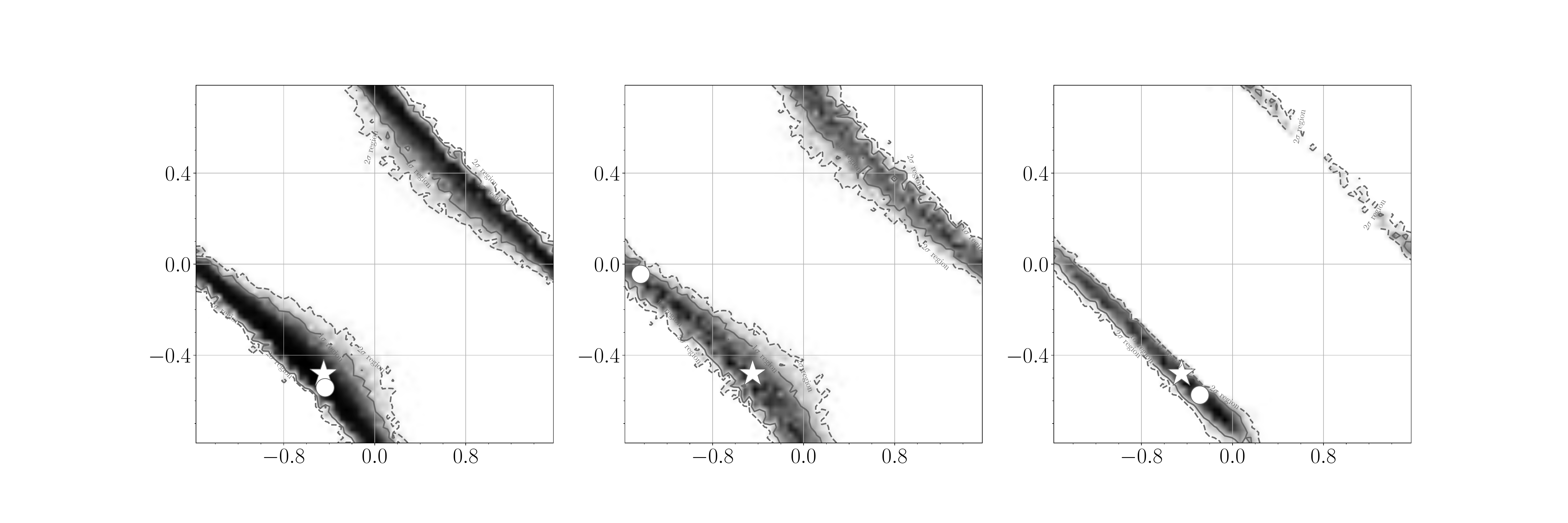}
 \includegraphics[width=16cm]{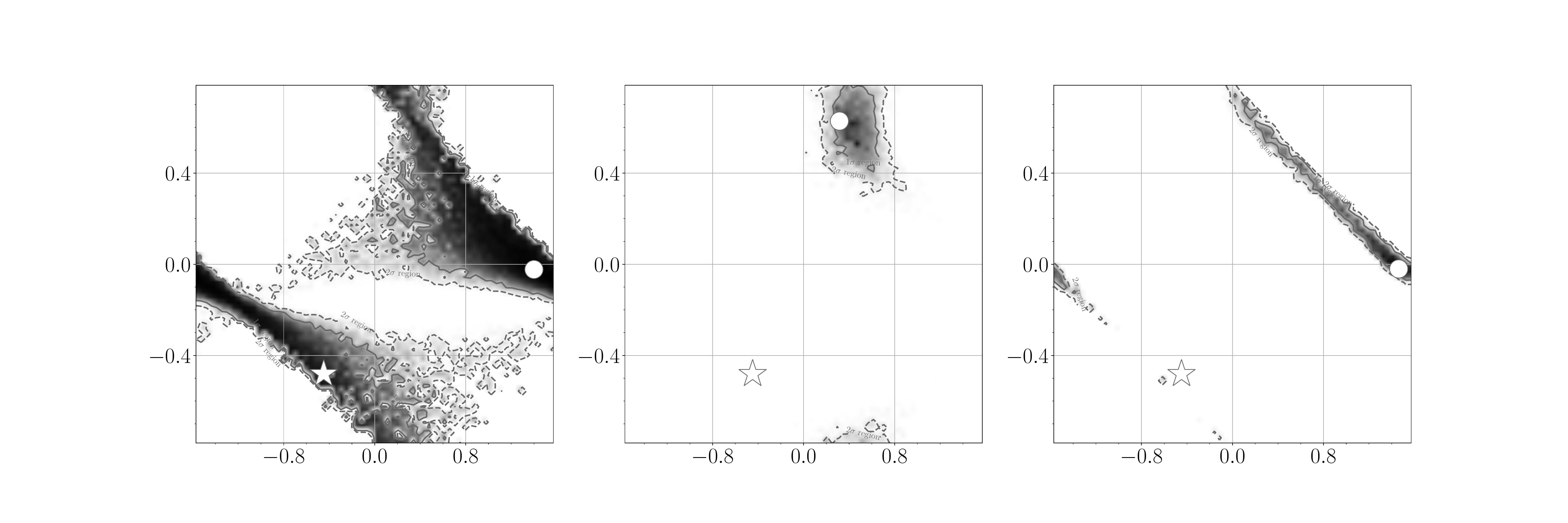}
    \includegraphics[width=16cm]{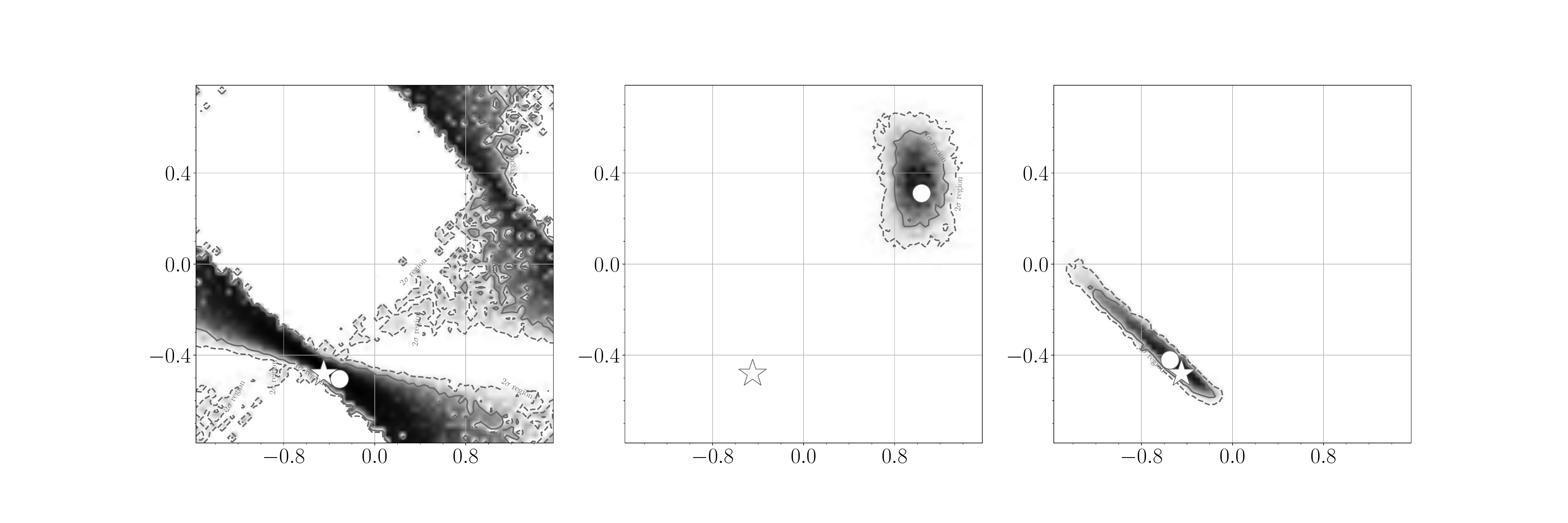}
 \includegraphics[width=16cm]{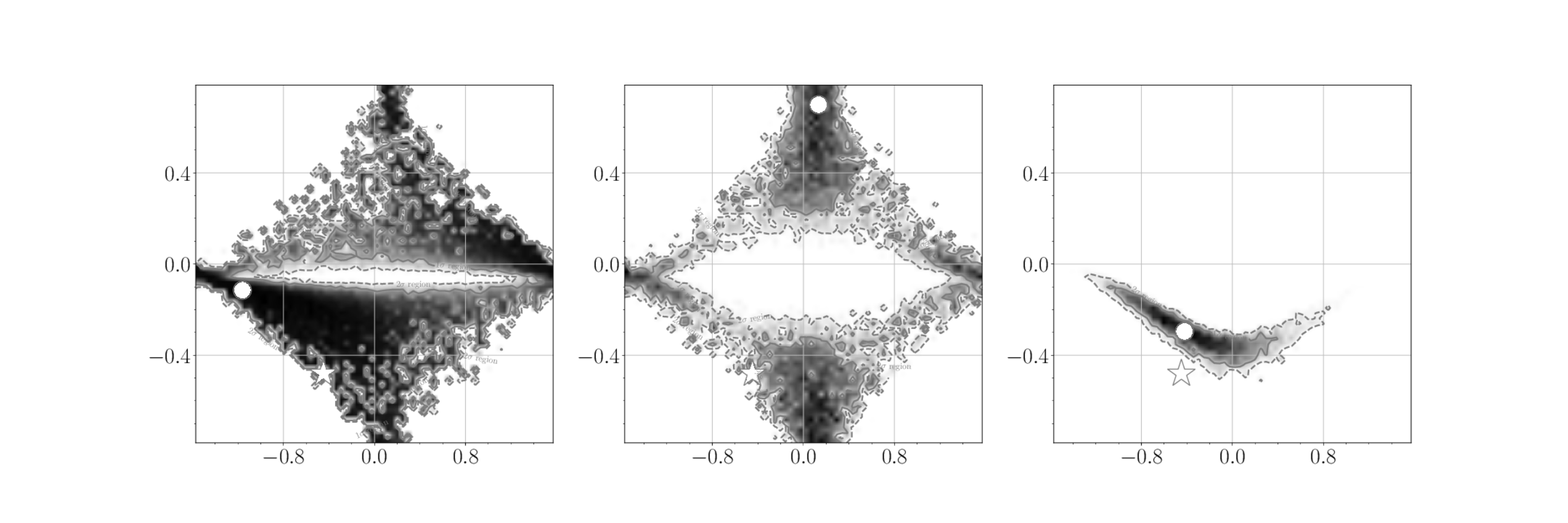}
    \caption{The probability distribution region represented by profile likelihood. The state 1 of the software-injected GW signal is used. The explanation of these figures are as same as Fig.~\ref{ken-short_thesis-fig:11}. The determinant values of beam-pattern function matrix are arranged $5\times10^{-4}, 1\times10^{-4}, 5\times10^{-5}, 1\times10^{-5}$ from top to bottom.\label{ken-short_thesis-fig:13}}
\end{figure*}

\begin{figure*}[tb]
  \centering
   \includegraphics[width=16cm]{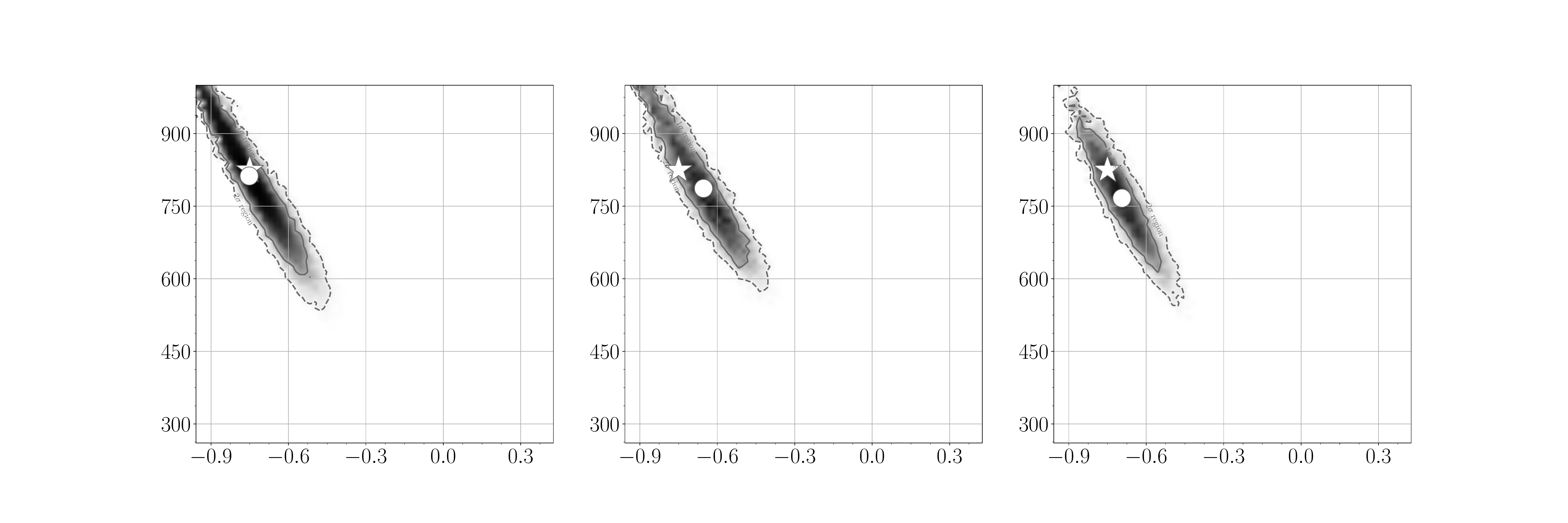}
 \includegraphics[width=16cm]{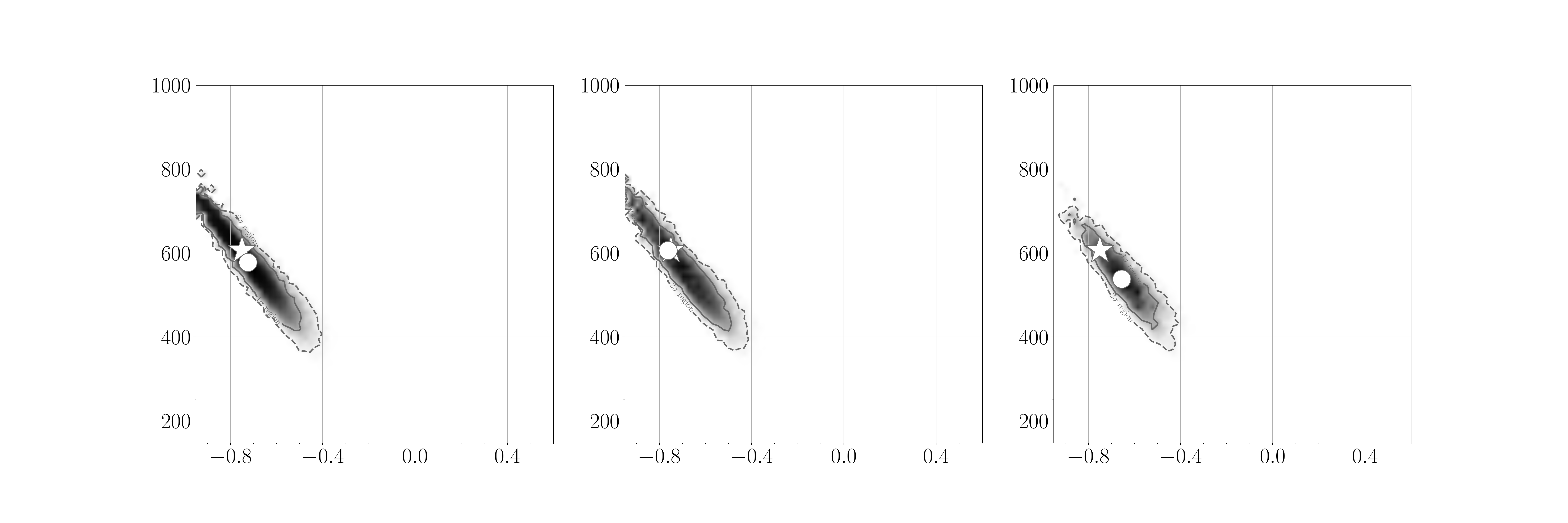}
    \includegraphics[width=16cm]{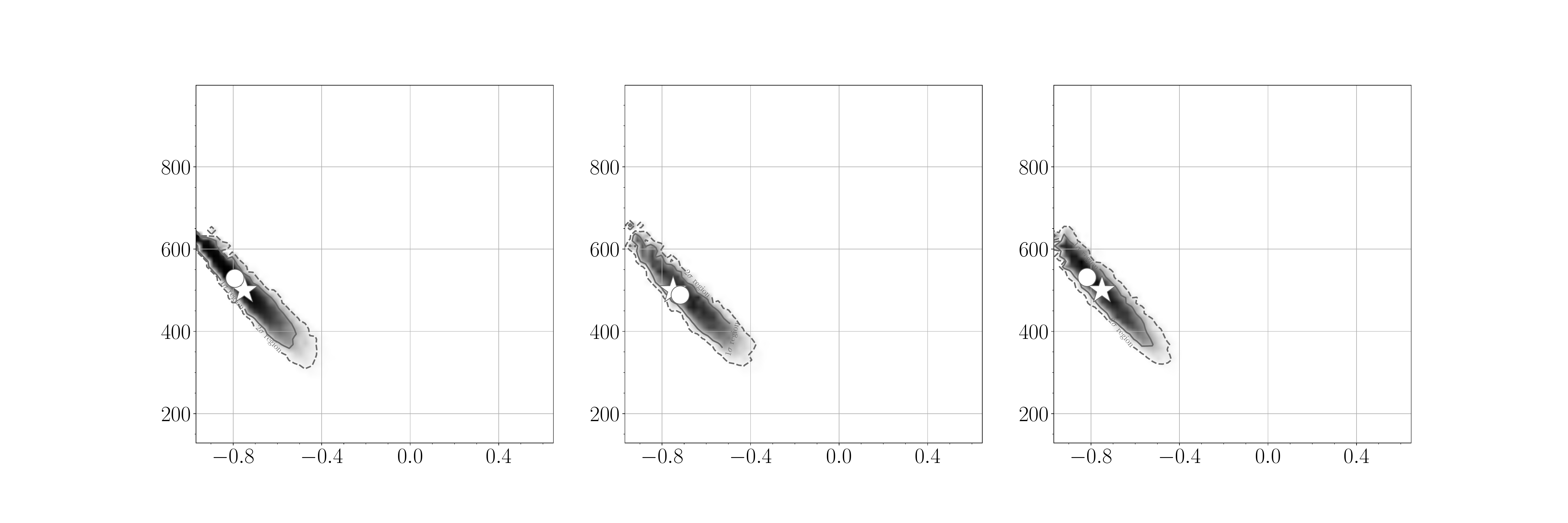}
   \caption{The probability distribution region represented by profile likelihood. The state 2 of the software-injected GW signal is used. The explanation of these figures are as same as Fig.~\ref{ken-short_thesis-fig:8}. The determinant values of beam-pattern function matrix are arranged $5\times10^{-1}, 1\times10^{-1}, 5\times10^{-2}$ from top to bottom.\label{ken-short_thesis-fig:20}}
\end{figure*}

\begin{figure*}[htbp]
 \centering
  \includegraphics[width=16cm]{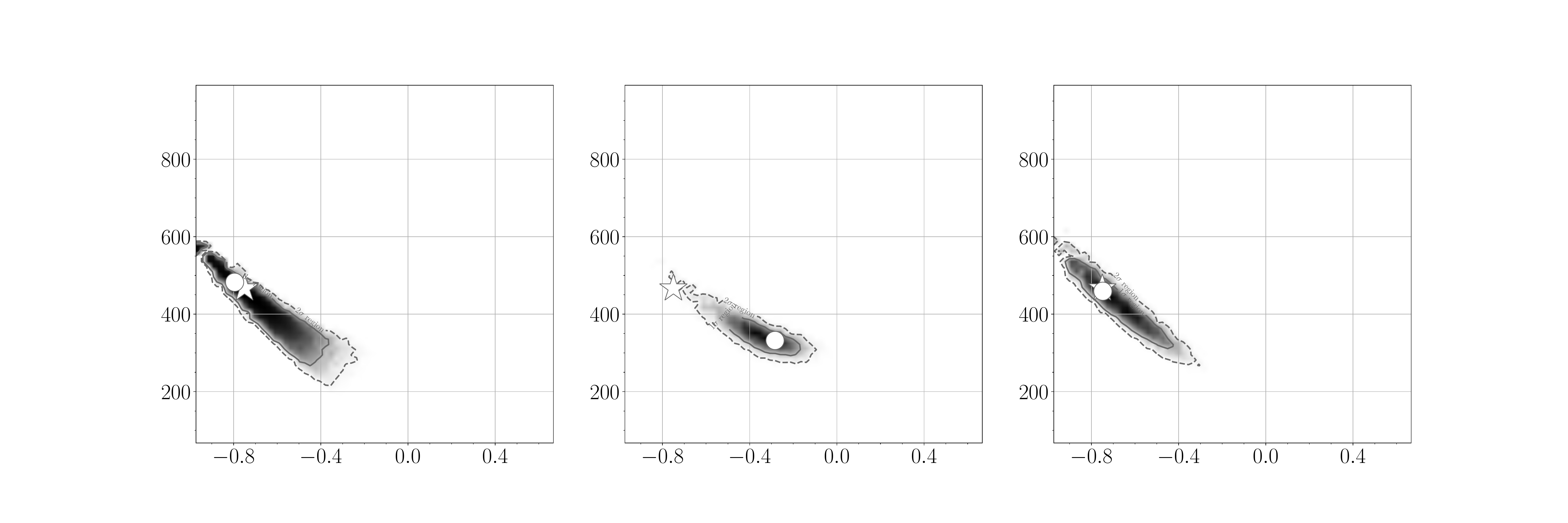}
    \includegraphics[width=16cm]{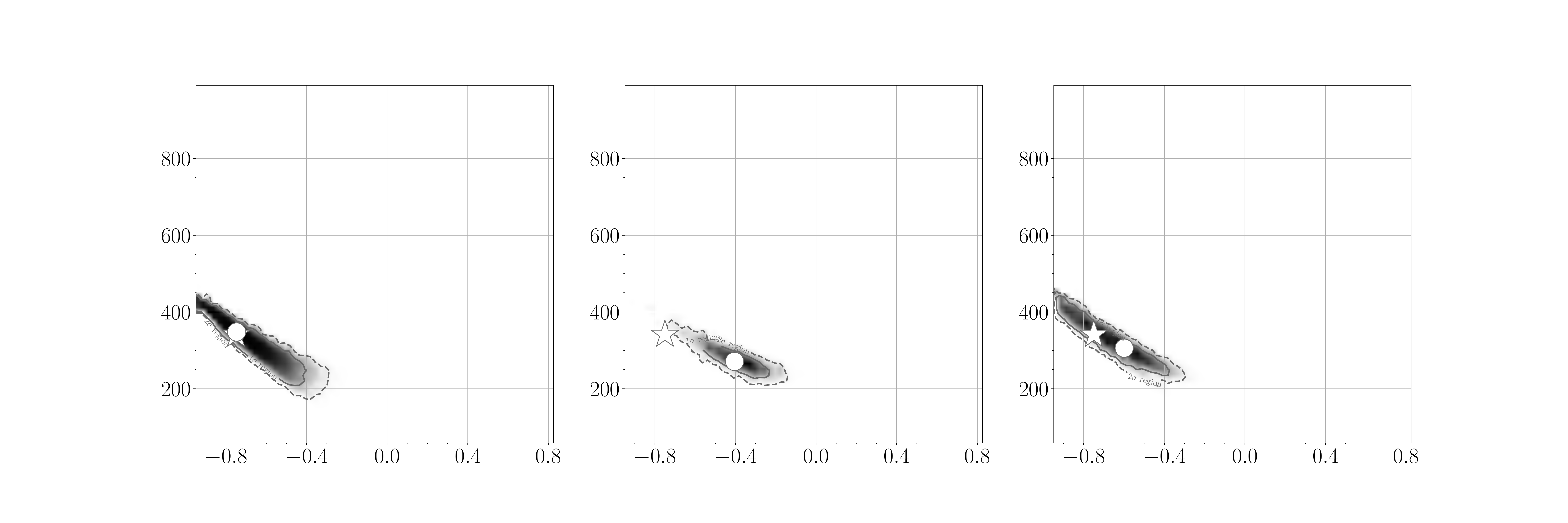}
 \includegraphics[width=16cm]{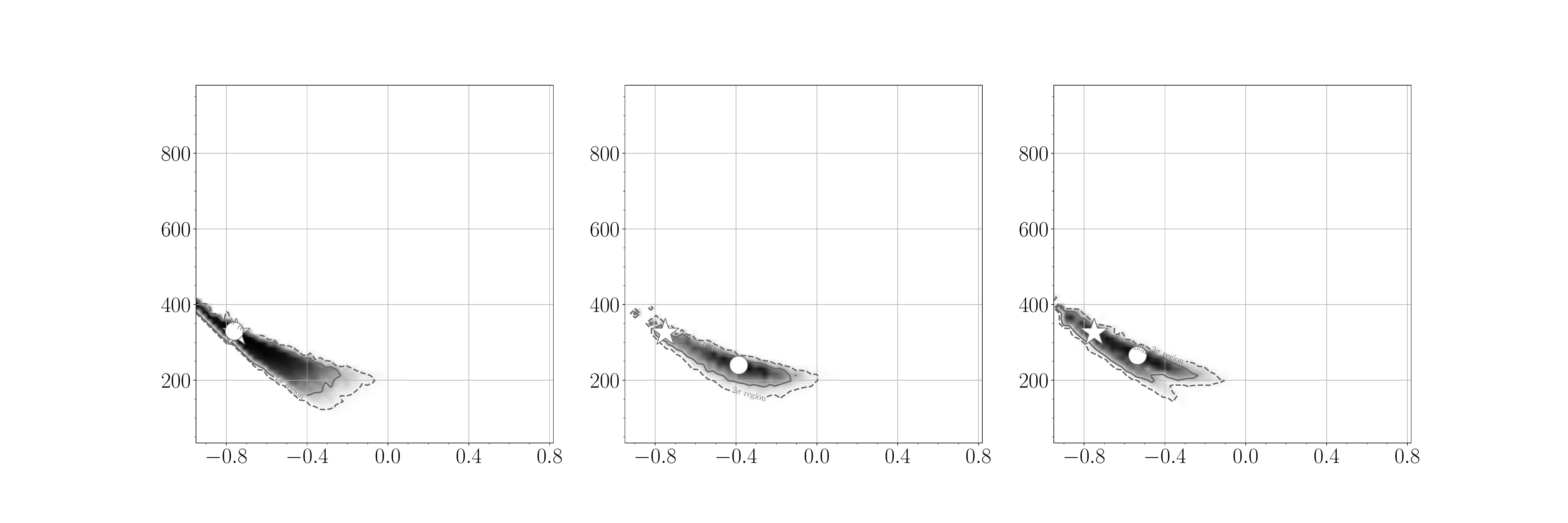}
    \caption{The probability distribution region represented by profile likelihood at the second data analysis. The explanation of these figures are as same as Fig.~\ref{ken-short_thesis-fig:8}. The determinant values of beam-pattern function matrix are arranged $1\times10^{-2}, 5\times10^{-3}, 1\times10^{-3}$ from top to bottom.\label{ken-short_thesis-fig:21}}
\end{figure*}

\begin{figure*}[htbp]
  \centering
    \includegraphics[width=16cm]{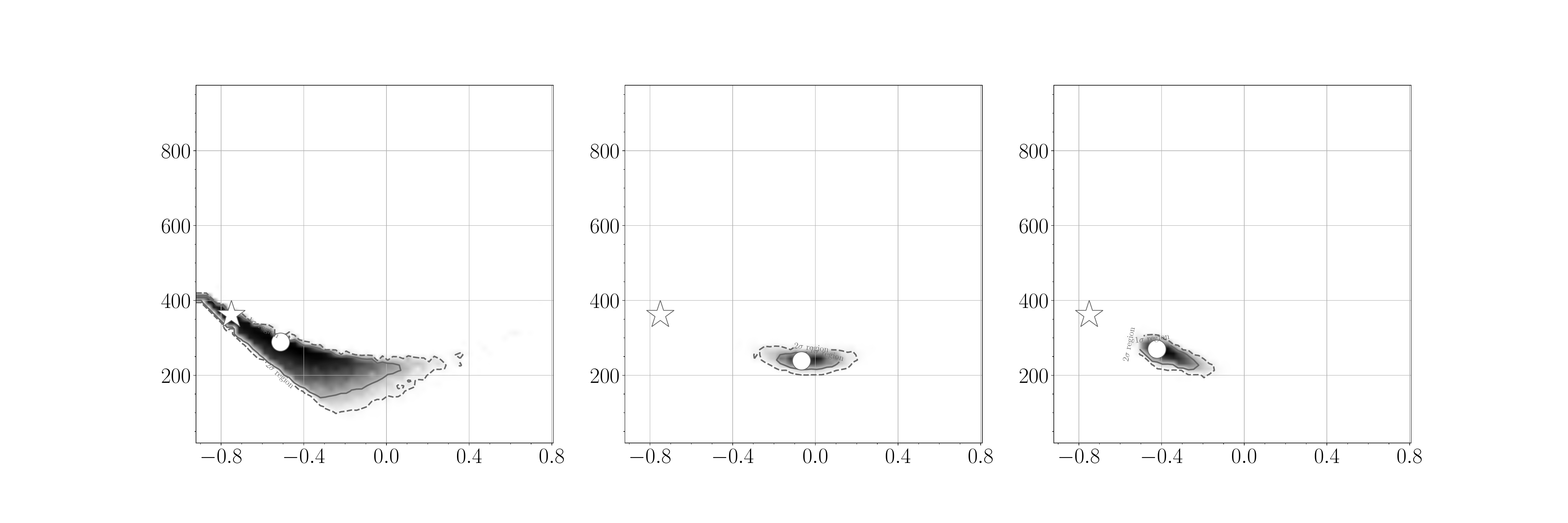}
 \includegraphics[width=16cm]{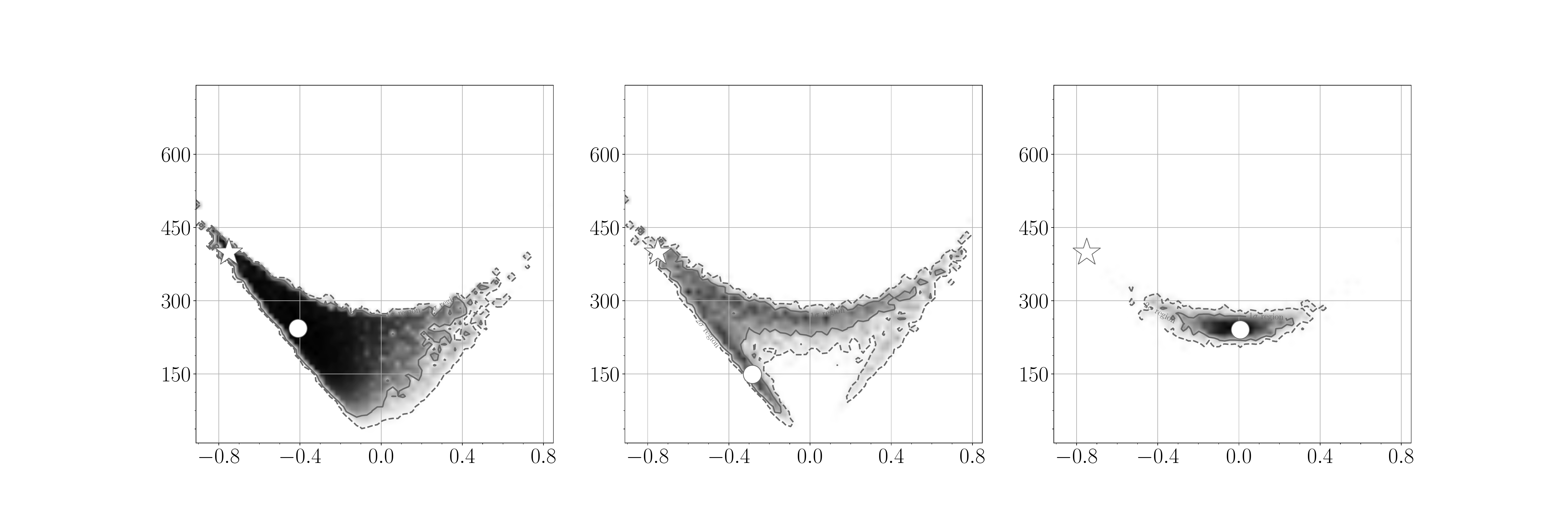}
    \includegraphics[width=16cm]{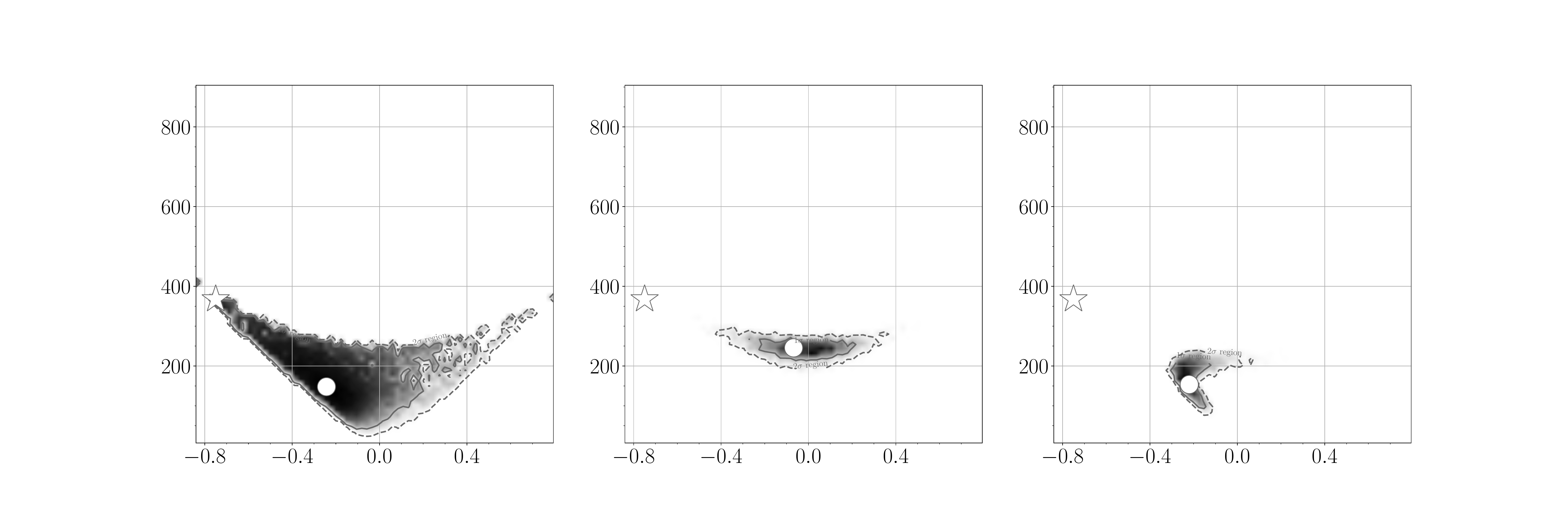}
 \includegraphics[width=16cm]{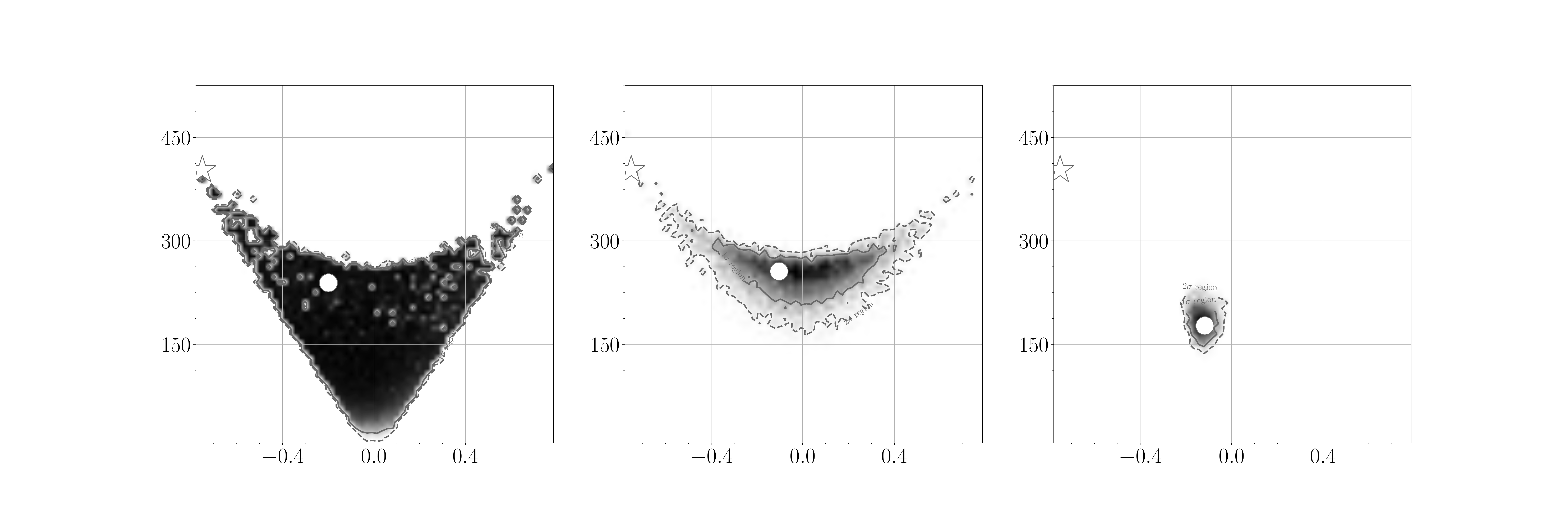}
    \caption{The probability distribution region represented by profile likelihood. The state 2 of the software-injected GW signal is used. The explanation of these figures are as same as Fig.~\ref{ken-short_thesis-fig:8}. The determinant values of beam-pattern function matrix are arranged $5\times10^{-4}, 1\times10^{-4}, 5\times10^{-5}, 1\times10^{-5}$ from top to bottom.\label{ken-short_thesis-fig:22}}
\end{figure*}

\begin{figure*}[tb]
  \centering
   \includegraphics[width=16cm]{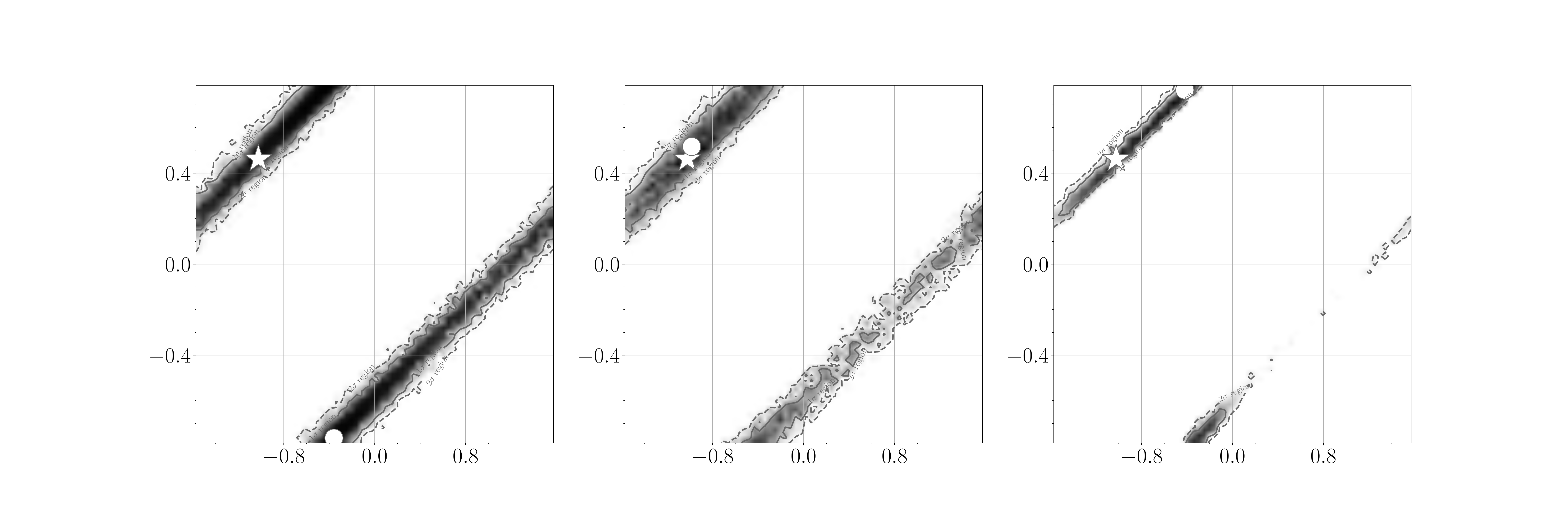}
 \includegraphics[width=16cm]{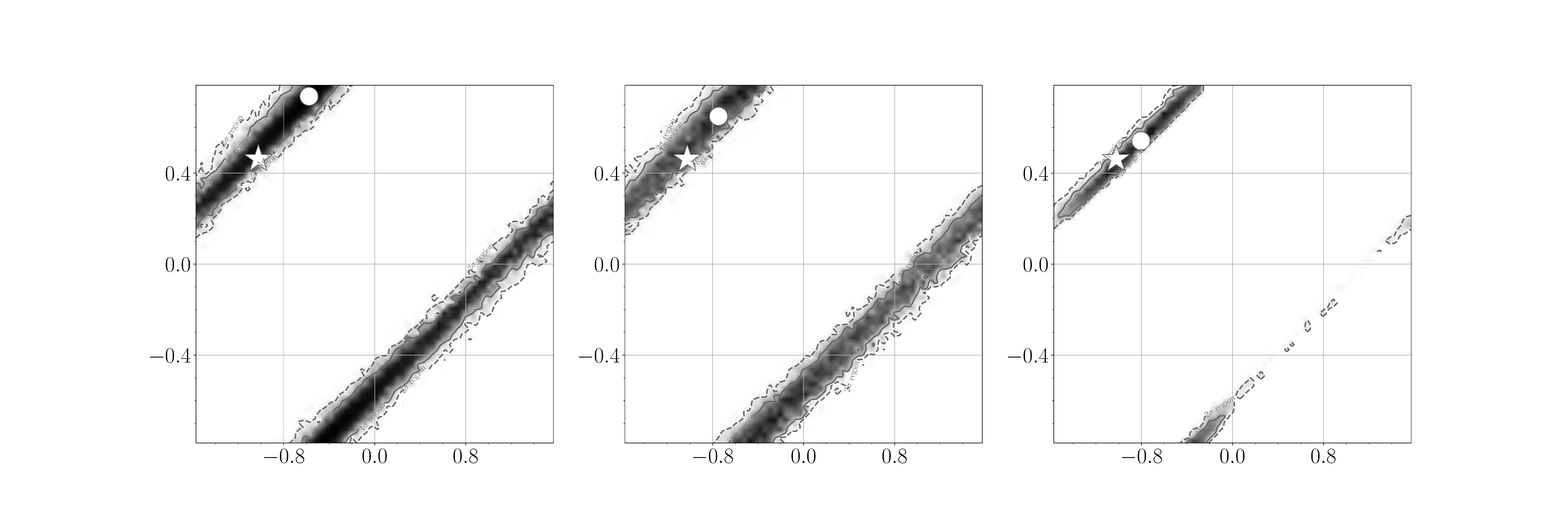}
    \includegraphics[width=16cm]{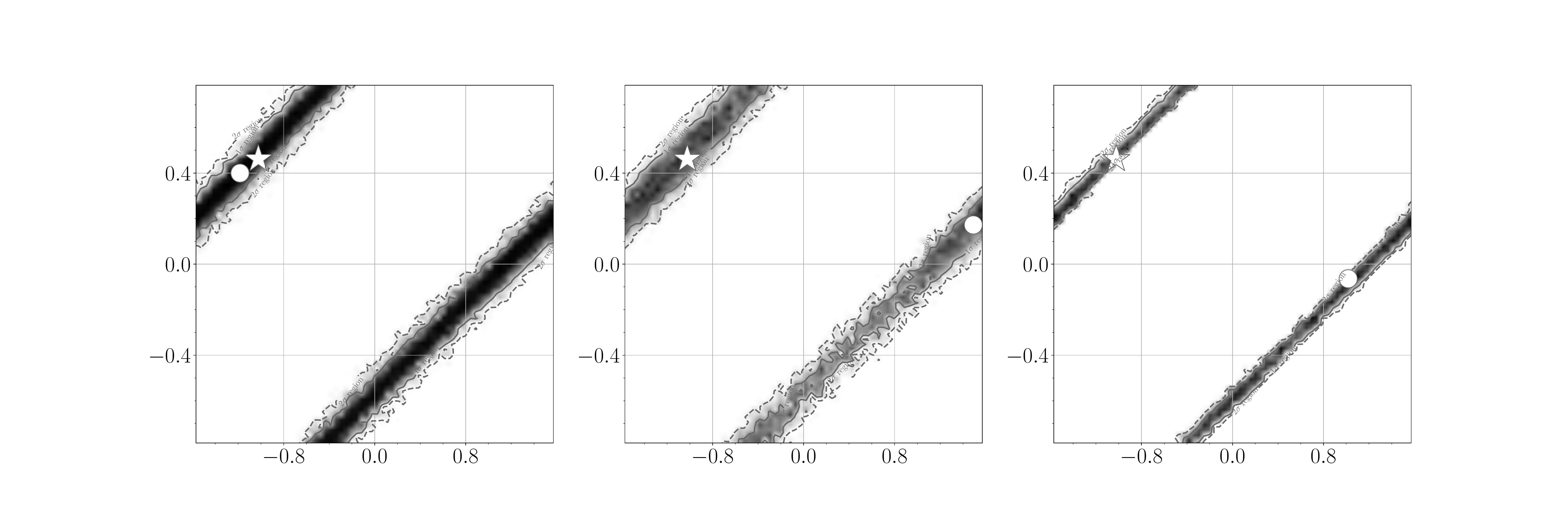}
   \caption{The probability distribution region represented by profile likelihood. The state 2 of the software-injected GW signal is used. The explanation of these figures are as same as Fig.~\ref{ken-short_thesis-fig:11}. The determinant values of beam-pattern function matrix are arranged $5\times10^{-1}, 1\times10^{-1}, 5\times10^{-2}$ from top to bottom.\label{ken-short_thesis-fig:23}}
\end{figure*}

\begin{figure*}[htbp]
 \centering
  \includegraphics[width=16cm]{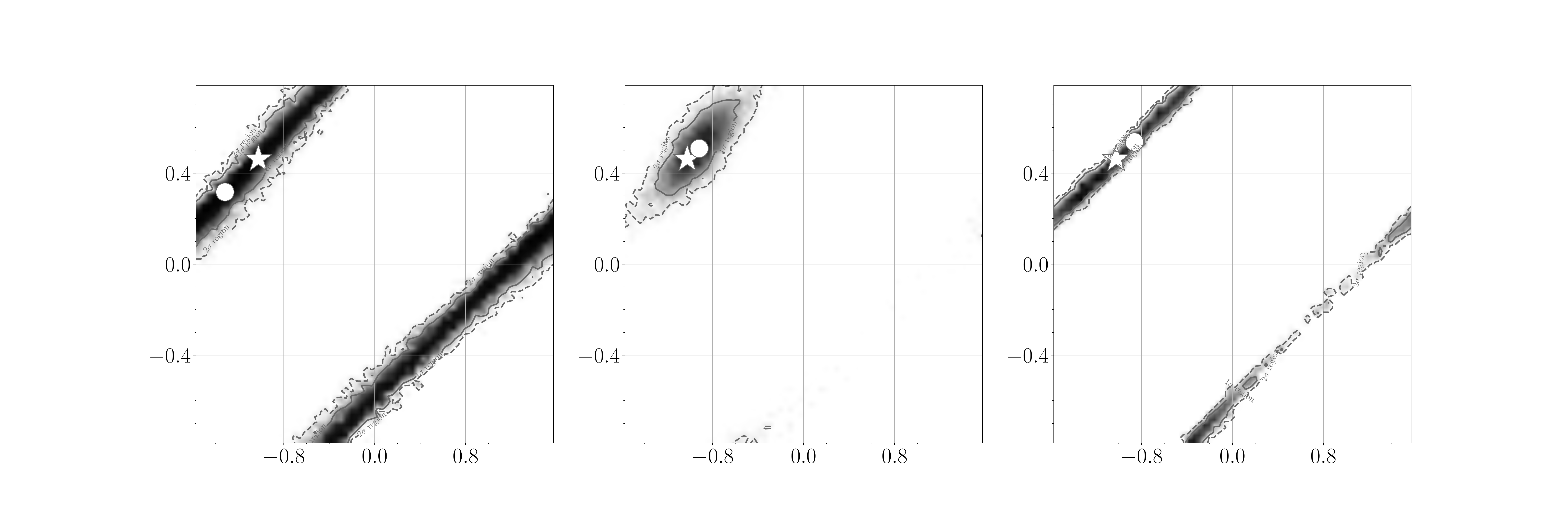}
    \includegraphics[width=16cm]{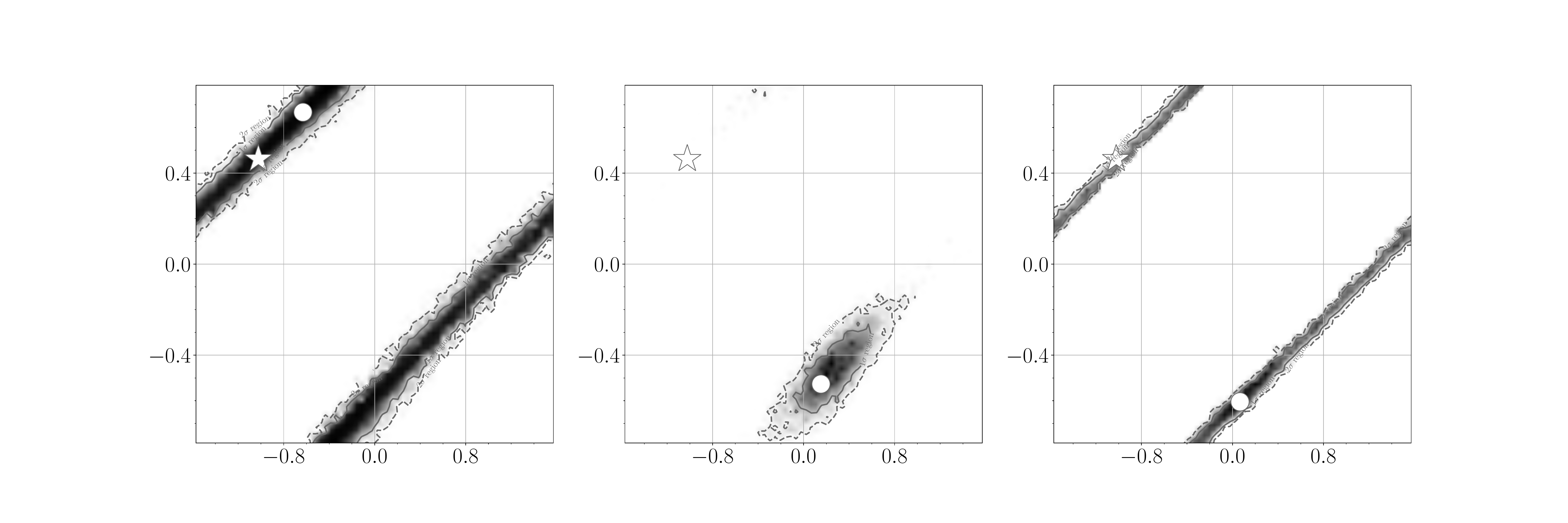}
 \includegraphics[width=16cm]{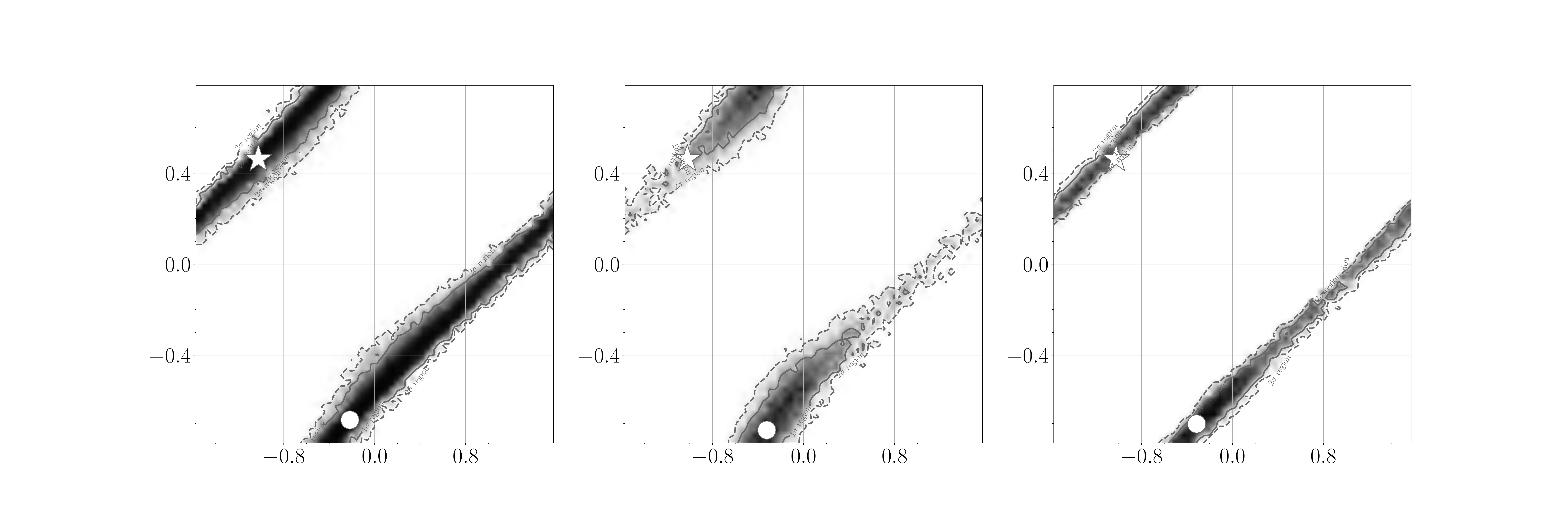}
    \caption{The probability distribution region represented by profile likelihood. The state 2 of the software-injected GW signal is used. The explanation of these figures are as same as Fig.~\ref{ken-short_thesis-fig:11}. The determinant values of beam-pattern function matrix are arranged $1\times10^{-2}, 5\times10^{-3}, 1\times10^{-3}$ from top to bottom.\label{ken-short_thesis-fig:24}}
\end{figure*}

\begin{figure*}[htbp]
  \centering
    \includegraphics[width=16cm]{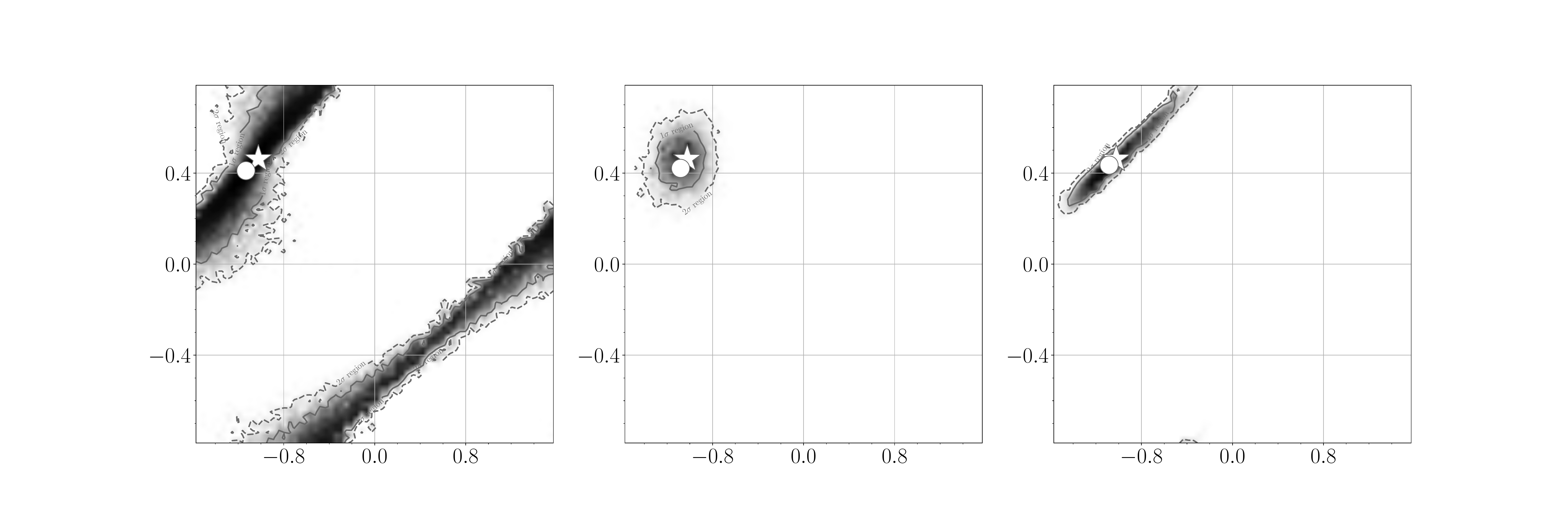}
 \includegraphics[width=16cm]{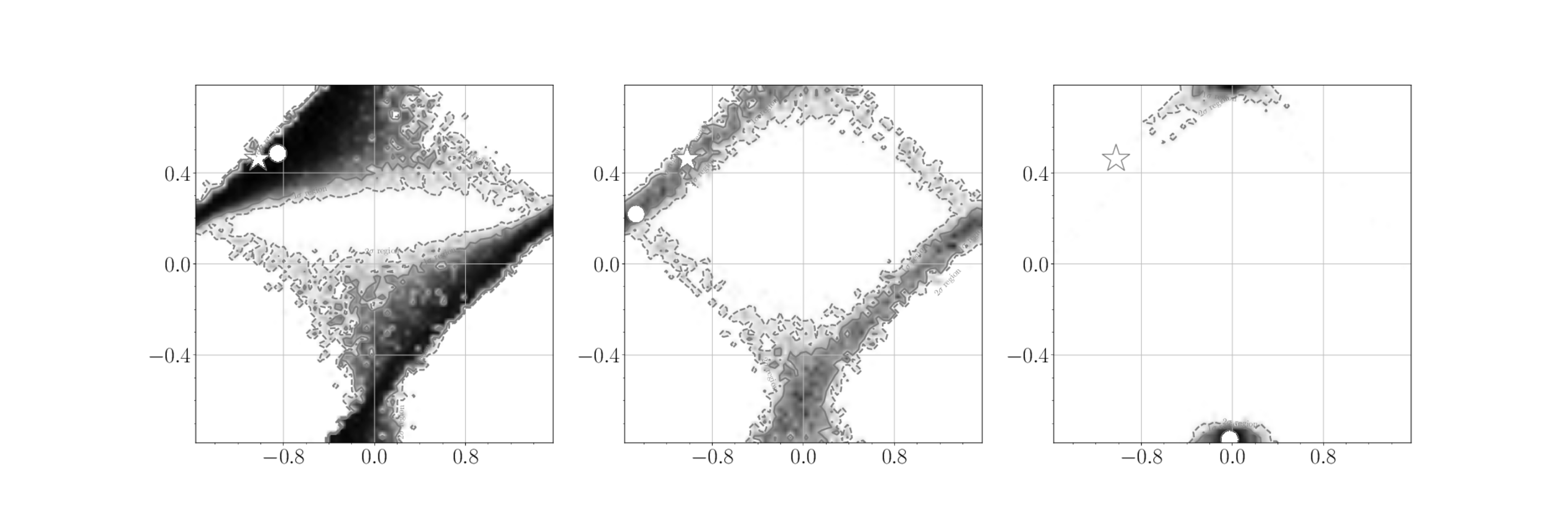}
    \includegraphics[width=16cm]{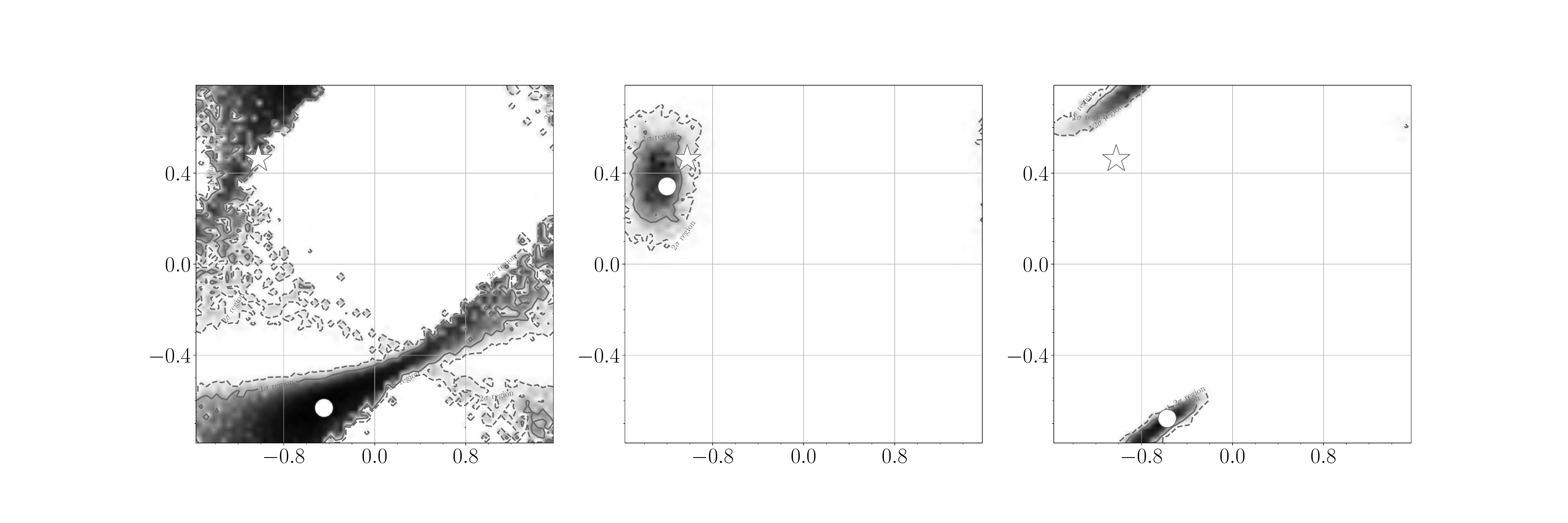}
 \includegraphics[width=16cm]{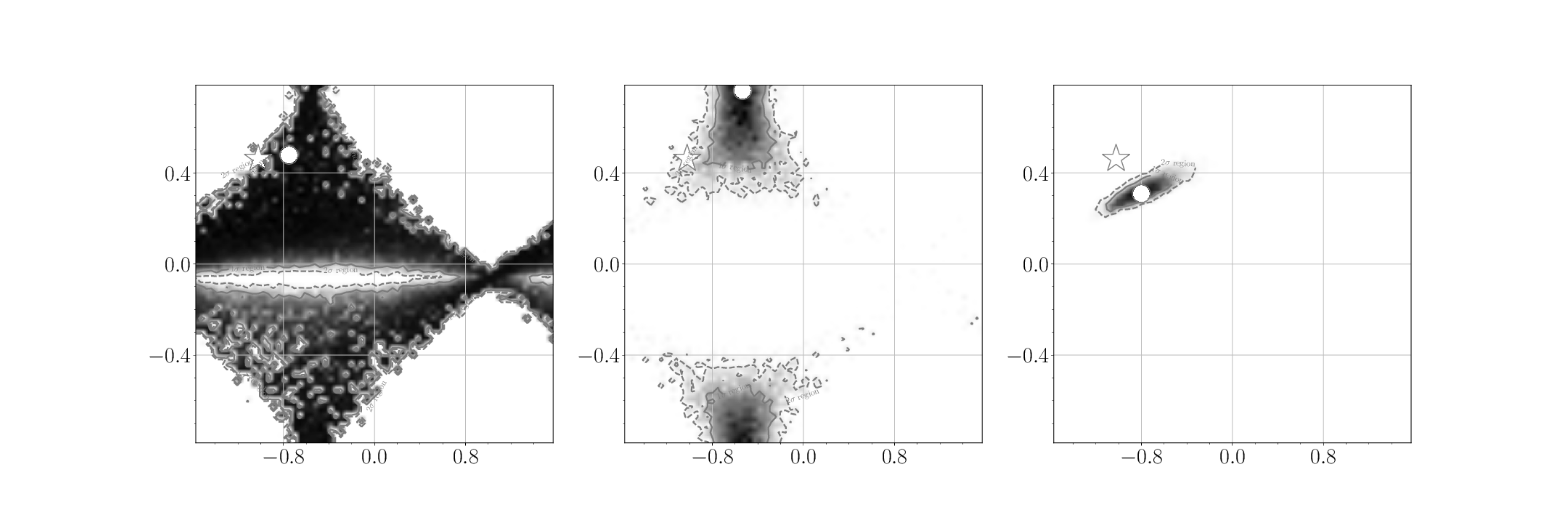}
    \caption{The probability distribution region represented by profile likelihood. The state 2 of the software-injected GW signal is used. The explanation of these figures are as same as Fig.~\ref{ken-short_thesis-fig:11}. The determinant values of beam-pattern function matrix are arranged $5\times10^{-4}, 1\times10^{-4}, 5\times10^{-5}, 1\times10^{-5}$ from top to bottom.\label{ken-short_thesis-fig:25}}
\end{figure*}

The data analysis results are expressed as the joint-probability distribution of the posterior as represented by the profile likelihood. The joint-probability distributions for all data analyses are shown in Figs.~\ref{ken-short_thesis-fig:8} - Fig.~\ref{ken-short_thesis-fig:25}. In these figures, the area enclosed by the bold line is the
$1\sigma$ (67\%) credible region and the dashed line is the
$2\sigma$ (95\%) credible region. The shrinkage rate of the credible region obtained by comparing the posterior distribution of the non-regularized and regularized data analysis using the type 1 and type 2 regulators is shown in Table.\ref{ken-short_thesis-tab:4} and \ref{ken-short_thesis-tab:5}, respectively. Here, state 1 and 2 of the software-injected data were used Table.\ref{ken-short_thesis-tab:4} and \ref{ken-short_thesis-tab:5}, respectively. An average of 100000 posterior samples for the non-regularized data analysis and 200000 posterior samples for regularized data analysis were evaluated. The implementation time of the algorithm was approximately one CPU hour for the non-regularized data analysis and approximately 50 CPU hour for the regularized data analysis.

The results clearly indicate that the accuracy of the amplitude parameters was improved by the regularized data analysis with the type 2 regulator in the most of the sky region; however, the accuracy of the amplitude parameters was primarily unimproved by the regularized data analysis with the type 1 regulator. Here, most actual values of the amplitude parameters were outside the credible region of the joint-posterior distribution. The average shrinkage rate of the credible region for the inclination vs luminosity distance all over the sky was approximately 1.5 times and that of the polarization vs initial phase was approximately 3.0 times for the regularized data analysis with the type 2 regulator. The shrinkage rate tends to be greater depending on the strength of the influence of the ill-posed inverse operator. However, in some sky region where the determinant value of the inverse matrix was greater than a certain threshold, the actual value was outside the credible region estimated by the regularized data analysis with the type 2 regulator. Our data analysis indicates the threshold of the determinant value depends on the actual inclination values of the CBC(i.e., the source of the GW signal). 

Here, we discuss the results of the data analysis for states 1 and 2 in detail. For the state 1, the regularized data analysis with the type 2 regulator improved the accuracy of the amplitude parameters in most of the sky region. In contrast, the actual value was outside the estimated credible region because the influence of the ill-posed inverse operator was significant in this case. This failure implies that the range of application of the regularized data analysis, where valid estimation can be implemented, is limited depending on the degree of influence the ill-posed inverse operator, even if the restriction in Eq.(\ref{ken-thesis-eq:51}) is imposed to prevent the credible region from lying outside of the actual parameter values. Fortunately, the sky region where the misestimation occurred is small and does not affect data analysis result significantly. The sky region wherein such a misestimation occurred accounts for
$<1\%$ for all over the sky in the case of state 1. Note that the significant reduction to the accuracy of luminosity distance, which is an important GW observation parameter, was realized at the determinant value, i.e., less than approximately $1\times10^{-3}$. This result allows us to treat a GW signal as a more precise standard siren\cite{schutz-nature}, and we can investigate cosmology without depending on the effect of bias from the location of the CBC source. In contrast to state 1, the result of the regularized data analysis for state 2 exhibits specific characteristics. The nature of the reduction of the accuracy of the amplitude parameters via the regularized data analysis is similar to the data analysis of state 1. However, the actual amplitude parameter values are outside the credible region of the inclination vs luminosity distance when the determinant value of the inverse operator is less than $5\times10^{-4}$, which threshold to the misestimation is larger than state 1. The main difference between states 1 and 2 is the absolute value of inclination, i.e., the absolute inclination value of state 2 expressed in cosine is greater than state 1, which may be related to the difficulty associated with correct estimation of the inclination-distance when the absolute inclination converted by cosine is relatively large.

\begin{table*}[htb]
   \begin{tabular}{|l||r|r|r||r|r|r||r|r|r||r|r|r|} \hline
    \multicolumn{1}{|l||}{} & \multicolumn{6}{c||}{Inclination vs luminosity distance} & \multicolumn{6}{c|}{Polarization vs initial phase}\\ \hline
    det & type 1 1$\sigma$ & type 1 2$\sigma$ &y/n &type 2 1$\sigma$ & type 2 2$\sigma$ &y/n &type 1 1$\sigma$ & type 1 2$\sigma$ &y/n & type 2 1$\sigma$ &type 2 2$\sigma$&y/n \\ \hline \hline
   $5\times10^{-1}$ & 0.94 & 1.00 &y &1.46 & 1.38 &y &1.02 & 0.92&y & 3.71 & 3.00&y \\ \hline
   $1\times10^{-1}$ & 1.09 & 0.95 &y &1.78 & 1.64 &y &1.29 & 1.09&y & 3.07 & 2.42&y \\ \hline
   $5\times10^{-2}$ & 0.77 & 0.82 &y &1.43 & 1.44 &y &0.88 & 0.80&y & 2.68 & 2.17&y \\ \hline
   $1\times10^{-2}$ & 1.35 & 1.25 &y &1.43 & 1.36 &y &1.43 & 1.49&y & 3.05 & 2.47&y \\ \hline
   $5\times10^{-3}$ & 1.03 & 1.26 &y &1.81 & 2.51 &y &1.28 & 1.45&y & 4.64 & 4.30&y \\ \hline
   $1\times10^{-3}$ & 1.37 & 1.90 &y &3.10 & 3.26 &y &1.66 & 2.13&y & 4.35 & 3.19&y \\ \hline
   $5\times10^{-4}$ & 1.06 & 1.12 &y &1.36 & 1.59 &y &1.04 & 1.15&y & 2.97 & 3.49&y \\ \hline
   $1\times10^{-4}$ & 4.11 & 7.73 &n &3.18 & 4.24 &y &5.90 & 9.25&n & 8.60 & 9.51&n \\ \hline
   $5\times10^{-5}$ & 3.95 & 6.57 &n &7.07 & 10.2 &y &4.72 & 7.36&n & 10.4 & 12.5&y \\ \hline
   $1\times10^{-5}$ & 0.96 & 1.29 &y &3.55 & 6.59 &n &1.17 & 1.74&n & 6.80 & 11.1&n \\ \hline
   \end{tabular}
 \caption{Rate of shrinkage of credible region from the joint-posterior distribution and indication of whether the $2\sigma$ credible region contains the actual point of the amplitude parameters. State 1 is used for the software-injected GW signal. The left-most column shows the determinant value of the beam-pattern matrix corresponding to the software-injected data. y/n indicates whether the actual points of the amplitude parameters are inside the $2\sigma$ credible region or not.\label{ken-short_thesis-tab:4}}
\end{table*}

\begin{table*}[htb]
   \begin{tabular}{|l||r|r|r||r|r|r||r|r|r||r|r|r|} \hline
    \multicolumn{1}{|l||}{} & \multicolumn{6}{c||}{inclination vs luminosity distance} & \multicolumn{6}{c|}{polarization vs initial phase}\\ \hline
    det & type 1 1$\sigma$ & type 1 2$\sigma$ &y/n &type 2 1$\sigma$ & type 2 2$\sigma$ &y/n &type 1 1$\sigma$ & type 1 2$\sigma$ &y/n & type 2 1$\sigma$ &type 2 2$\sigma$&y/n \\ \hline \hline
    $5\times10^{-1}$ & 1.06 & 1.09&y & 1.23 & 1.50&y & 1.14 & 1.40&y & 4.04 & 4.48&y \\ \hline
    $1\times10^{-1}$ & 1.00 & 1.00&y & 1.18 & 1.52&y & 1.00 & 0.98&y & 3.92 & 4.30&y \\ \hline
    $5\times10^{-2}$ & 0.95 & 0.99&y & 1.01 & 1.05&y & 1.07 & 1.23&y & 2.78 & 2.41&y \\ \hline
    $1\times10^{-2}$ & 1.66 & 2.81&n & 1.36 & 1.57&y & 2.51 & 3.90&y & 3.14 & 3.48&y \\ \hline
    $5\times10^{-3}$ & 1.52 & 2.52&n & 1.11 & 1.15&y & 2.65 & 4.53&n & 2.75 & 4.54&y \\ \hline
    $1\times10^{-3}$ & 1.12 & 1.31&y & 1.19 & 1.27&y & 1.24 & 1.93&y & 1.97 & 2.10&y \\ \hline
    $5\times10^{-4}$ & 4.54 & 6.33&n & 5.11 & 5.97&n & 5.10 & 7.80&y & 8.15 & 8.69&y \\ \hline
    $1\times10^{-4}$ & 1.20 & 1.83&y & 4.06 & 6.67&n & 1.63 & 2.35&y & 11.5 & 19.5&n \\ \hline
    $5\times10^{-5}$ & 3.96 & 7.85&n & 6.69 & 10.1&n & 4.68 & 7.28&y & 9.96 & 12.3&n \\ \hline
    $1\times10^{-5}$ & 1.92 & 4.48&n & 13.6 & 26.7&n & 3.04 & 6.60&y & 24.5 & 44.9&n \\ \hline
   \end{tabular}
 \caption{Rate of shrinkage of credible region from the joint-posterior distribution and indication of whether the $2\sigma$ credible region contains the actual point of the amplitude parameters. State 1 is used for the software-injected GW signal. State 2 was used for the software-injected GW signal.\label{ken-short_thesis-tab:5}}
\end{table*}

\section{Conclusion}\label{server-thesis-sec:conclusion}
In this study, we discuss the method to prevent the accuracy of amplitude parameters of a GW from a CBC from being amplified owing to the ill-posed nature of the inverse operator comprising the beam-pattern functions of a network of GW telescopes. The key idea to resolve this problem is compensation via an appropriate correction term represented as a matrix called a ``regulator'' for an inverse operator whose rank is deficient. The Tikhonov type regulator(Eq.(\ref{ken-short_thesis-eq:1}), type 1) and symmetric trace-free type regulator regulator(Eq.(\ref{ken-short_thesis-eq:2}), type 2) are used to express the regulator matrix. A regulator reduce the amplified noise caused by the ill-posed problem, whereas the value of the regulator affects the parameter accuracy as bias noise. To resolve the bias problem, the optimization of the regulator is evaluated by minimizing the sum of these noises. A Lagrange multiplier method with KKT condition for the norm of the difference between the amplitude parameters estimated by the regularized data analysis and amplitude parameters estimated by the non-regularization method (Eq.(\ref{ken-thesis-eq:44})), i.e., the a-posteriori parameter choice rule provides optimized regulator values. The data analysis for two types of software-injected GW signal using the MultiNest software was implemented to evaluate the reduction accuracy of the amplitude parameters using proposed method. The data analysis results indicate that the regularization method with the type 2 regulator reduces the credible region of the accuracy of the amplitude parameters. For approximately 90\% of the sky region, the credible region of inclination-distance is reduced by approximately 1.5 times and that of the polarization-initial phase is reduced by approximately 3.0 times. The shrinkage rate of the credible region increases with a decreasing determinant value of the inverse operator; however the actual value lies outside the credible region in the regularized data analysis owing to the significant ill-posed inverse operator. The threshold to maintain the validity of the regularization method appears to be related the inclination values.

We consider that the proposed method, which employ a regularization method for a targeted GW signal from a CBC and uses an optimized regulator, represents robust and coherent data analysis that enables us to minimize the amplified noise caused by an ill-posed nature of the inverse operator. In other words, for most of the sky region, it is possible that the accuracy of the amplitude parameters estimated by the proposed method is nearly the same as the limit of the accuracy as calculated by the SNR of the signal. Thus, the proposed method is suitable for the parameter estimation of a CBC search.

\begin{acknowledgments}
We thank Masatake Ohashi for helpful comments. This study was supported by MEXT, JSPS Leading-edge Research Infrastructure Program, JSPS Grant-in-Aid for Specially Promoted Research 26000005, JSPS Core-to-Core Program, Advanced Research Networks, the joint research program of the Institute for Cosmic Ray Research, Grant-in-Aid for Scientific Research on Innovative areas (No.2905, No.17H06357, No.17H06365).
\end{acknowledgments}

\nocite{*}
% \bibliographystype{apsrev4-1}
\bibliographystyle{apsrev4-1}
\bibliography{thesis} % Produces the bibliography via BibTeX.

\end{document}